\documentclass[times,twocolumn,final]{elsarticle}

\usepackage{medima}
\usepackage{framed,multirow}

\usepackage{amssymb}
\usepackage{latexsym}

\usepackage{graphicx}
\usepackage[colorlinks,allcolors=black,pdftex]{hyperref}
\usepackage{amsmath}
\usepackage[table,dvipsnames]{xcolor}
\usepackage{subcaption}
\usepackage{float}
\usepackage{bbding}
\usepackage{array}
\usepackage[parse-numbers=false]{siunitx}
\usepackage{multicol}
\usepackage{multirow}
\usepackage{threeparttable}
\usepackage{booktabs}
\usepackage{rotating}
\newcolumntype{P}[1]{>{\centering\arraybackslash}p{#1}}
\newcolumntype{M}[1]{>{\centering\arraybackslash}m{#1}}
\usepackage{url}
\usepackage{soul}
\usepackage{hyperref}
\usepackage[toc,page]{appendix}
\usepackage{subfiles}
\usepackage{natbib}

\definecolor{newcolor}{rgb}{.8,.349,.1}

\journal{Medical Image Analysis}

\begin{document}

\verso{Mojtaba Lashgari \textit{et~al.}}

\begin{frontmatter}

\title{Three-dimensional micro-structurally informed in silico myocardium-- towards virtual imaging trials in cardiac diffusion weighted MRI}%

\author[1,2]{Mojtaba \snm{Lashgari}}\corref{}
\ead{m.lashgari@leeds.ac.uk}
\cortext[]{Corresponding author: Center for Computational Imaging and Simulation Technologies in Biomedicine (CISTIB), School of Computing, University of Leeds, Leeds, UK.}
\author[1,2]{Nishant \snm{Ravikumar}}
\author[2]{Irvin \snm{Teh}}
\author[3]{Jing-Rebecca \snm{Li}}
\author[2]{David \snm{L.Buckley}}
\author[2]{Jurgen \snm{E. Schneider}}
\author[1,2,3,5,6]{Alejandro \snm{F. Frangi}}\corref{}
\ead{a.frangi@leeds.ac.uk}

\address[1]{Centre for Computational Imaging and Simulation Technologies in Biomedicine (CISTIB), School of Computing, University of Leeds, Leeds, UK}
\address[2]{Biomedical Imaging Science Department, Leeds Institute for Cardiovascular and Metabolic Medicine (LICAMM), School of Medicine, University of Leeds, Leeds, UK}
\address[3]{INRIA Saclay, Equipe DEFI, CMAP, Ecole Polytechnique, Route de Saclay, 91128 Palaiseau Cedex, France}
\address[4]{Medical Imaging Research Center (MIRC), Department of Cardiovascular Sciences, KU Leuven, Leuven, Belgium.}
\address[5]{Medical Imaging Research Center (MIRC), Department of Electrical Engineering, KU Leuven, Leuven, Belgium.}
\address[6]{Alan Turing Institute, London, UK.}

\received{1 May 2013}
\finalform{10 May 2013}
\accepted{13 May 2013}
\availableonline{15 May 2013}
\communicated{S. Sarkar}

\begin{abstract}
In silico tissue models (viz. numerical phantoms) provide a mechanism for evaluating quantitative models of magnetic resonance imaging.
This includes the validation and sensitivity analysis of imaging biomarkers and tissue microstructure parameters.
This study proposes a novel method to generate a realistic numerical phantom of myocardial microstructure.
The proposed method extends previous studies by accounting for the variability of the cardiomyocyte shape, water exchange between the cardiomyocytes (intercalated discs), disorder class of myocardial microstructure, and four sheetlet orientations.
In the first stage of the method, cardiomyocytes and sheetlets are generated by considering the shape variability and intercalated discs in cardiomyocyte\textemdash cardiomyocyte connections.
Sheetlets are then aggregated and oriented in the directions of interest.
The morphometric study demonstrates no significant difference ($p>0.01$) between the distribution of volume, length, and primary and secondary axes of the numerical and real (literature) cardiomyocyte data.
Moreover, structural correlation analysis validates that the in-silico tissue is in the same class of disorderliness as the real tissue.
Additionally, the absolute angle differences between the simulated helical angle (HA) and input HA (reference value) of the cardiomyocytes ($4.3^\circ\pm3.1^\circ$) demonstrate a good agreement with the absolute angle difference between the measured HA using experimental cardiac diffusion tensor imaging (cDTI) and histology (reference value) reported by (Holmes et al., 2000) ($3.7^\circ\pm6.4^\circ$) and (Scollan et al., 1998) ($4.9^\circ\pm14.6^\circ$).
Furthermore, the angular distance between eigenvectors and sheetlet angles of the input and simulated cDTI is much smaller than those between measured angles using structural tensor imaging (as a gold standard) and experimental cDTI.
Combined with the qualitative results, these results confirm that the proposed method can generate richer numerical phantoms for the myocardium than previous studies.
\end{abstract}


\end{frontmatter}


\section{Introduction}
Cardiovascular diseases (CVDs) are a significant global health concern, accounting for more than one-quarter of all global deaths each year ($\sim$ 17.5 million) \citep{ezzati2015contributions}. 
Physical changes in the microstructure of the myocardium accompany many CVDs, like cardiomyopathies. 
Diffusion magnetic resonance imaging (dMRI) is a physical measurement of the stochastic motion of water molecules.
During a typical magnetic resonance diffusion measurement, i.e., with diffusion times of $10$\textemdash\SI{100}{\milli\second}, water molecules diffuse around $6$\textemdash$\SI{18}{\micro\meter}$ within the soft tissues with a diffusivity of $D=\SI{1.5}{\micro\meter^2\per\milli\second}$ \citep{yu2017tissue}. 
However, the motion of water molecules is hindered and restricted in the cellular environment. 
This creates the opportunity to investigate tissues at the microscale, enabling a more in-depth study of pathological processes. 
\par To understand the microstructural changes that accompany the onset and progression of different types of cardiac pathology, there has been an increasing interest in applying dMRI models to cardiology \citep{nielles2019cardiac,mekkaoui2017diffusion}.
Methods such as diffusion tensor imaging (DTI) and diffusion kurtosis imaging fit different signal models to the diffusion signal.
However, they do not provide biophysical interpretations of the dMRI signal, which would be of more direct clinical relevance.
A bottleneck for developing biophysical models is the lack of suitable phantoms for their validation. 
Biophysical models are formulated based on approximations and assumptions derived from the biological information of the underlying tissue structure. 
Both physical and numerical cardiac phantoms would be desirable to validate these assumptions and approximations.
\par The advantages of numerical phantoms over their physical counterparts are a reduction in the time and costs associated with complete an imaging study, reproducibility of the results \citep{sauer2022anatomically}, relative ease to control the model detail and complexity of the tissue microstructure, and prevention of the pronounced artefacts in the result of physical phantom due to dismissing properties of unknown material in the physical phantom \citep{fieremans2018physical}.
Numerical phantoms\textemdash using Monte Carlo (MC) or finite element method (FEM) based simulators \textemdash could be used to aid in biophysical modelling and interpretation of cardiac dMRI signals, design and conduct a feasibility study of the corresponding inverse problem to be solved, and analyse the sensitivity of dMRI measurements to changes in tissue microstructure and the use of different pulse sequences \citep{bates2017monte, rose2019novel}.
Moreover, integrating an accurate cardiac numerical phantom with a validated imaging simulator facilitates virtual imaging trials in cardiac dMRI.
The virtual imaging trial is a unique alternative to assess and optimise medical imaging technologies by imitating the virtual imaging trials and studying the states that are not physically realisable nor ethically responsible in vivo \citep{abadi2021virtual, sauer2022anatomically, abadi2020virtual}.
In addition, virtual imaging trials promise to lead to faster, safer and cost-effective regulatory studies compared to conventional clinical trials, due to replacing actual patients and imaging devices with virtual surrogates \citep{abadi2020virtual}. 
Nevertheless, few studies have explored the benefits and challenges associated with designing numerical phantoms of the myocardium.
The first cardiac numerical phantom was proposed by \citep{wang2011simulation}.
They generated a multi-scale numerical phantom across numerous spatial resolutions, and used these phantoms to produce dMRI signal via MC simulations.
Their numerical phantom included a packed arrangement of cardiomyocytes (CMs), simplified as cylindrical geometries with hexagonal cross sections. 
In a subsequent study, the same authors presented a different phantom design, modelling CMs as cylinders with different sizes, arrangements, and length/diameter ratios with varying numbers of CMs ($1, 8, 64$, and $8000$). 
Next, they modelled sheetlet structures by arranging CMs based on their orientation \citep{wang2012multiscale}.
We previously introduced a numerical phantom in which the length, cross-sectional area, and aspect ratio (thickness/width) of CMs were based on data from the literature \citep{bates2017monte}. 
In our approach, the CMs were modelled as rectangular cuboids and arranged in two stages. 
The CMs were first arranged parallelly to form a sheetlet (sub-voxel stage). 
Next, adjacent sheetlets within a voxel were progressively rotated to simulate the transmural variation in the \textit{helix angle} (HA) (sheetlet or voxel stage) \citep{bates2017monte}. 
In a recent study, \citep{rose2019novel} generated a numerical phantom based on histological images of the heart. 
CMs were first manually segmented from a sub-voxel area of the image, and the segmentation boundaries were then approximated with polygons with an average of 99 vertices. 
Next, by extruding individual CMs along the normal perpendicular to their boundary in the 2D plane, a 3D block of CMs was created with random uniformly distributed lengths. 
Finally, they modelled HA by placing several blocks next to each other using different angles for the longitudinal axis of the cardiac local coordinate system \citep{rose2019novel}.

The goal of this study is to address the limitations of the previous numerical phantoms of the heart muscle tissue, namely the lack of a detailed numerical phantom and the failure to incorporate the cellular shape variability. 
Concerning the first limitation, numerical phantom mimics the most relevant biophysical characteristics of extracellular space (ECS) and intracellular space (ICS) within the myocardium.
There is evidence to suggest that regional curvature of ventricles \citep{su2012geometrical,weisman1985global,espe2017regional}, collagen in ECS matrix \citep{haddad2017novel}, intercalated discs (ICDs) in ICS \citep{pinali2015three, perriard2003dilated, noorman2009cardiac}, and changes in their respective biophysical characteristics are closely linked to some CVDs.
Although these changes may affect diffusion-weighted (DW) signals, no study to date has modelled these characteristics and focused solely on modelling the CMs and ECS.
Concerning the second limitation, most studies typically consider identical sets of simple geometries for cellular shape, limiting the complexity of the myocardium captured by numerical phantoms. 
These phantoms perform poorly compared with simulations based on phantoms generated from histological images of tissues \citep{naughton2019connecting}.
Moreover, it has been demonstrated that the disorder class of microstructure (determined based on the cellular shape variability and packing disorderliness) affects the coefficients of the time dependence diffusion at the macroscale \citep{novikov2014revealing}. 
Consequently, it is imperative to consider the native probability distribution functions (PDFs) of the biophysical parameters associated with CMs shape (introduced in the next Section) when devising a realistic numerical phantom.
Finally, we investigate the use of FE analysis to simulate the dMRI signal of the myocardium, which is more accurate than the conventional MC approach \citep{grebenkov2016microstructure}. 
To the best of our knowledge, this has not been explored previously in the heart.


\section{Relevant Myocardial Tissue Properties}
This section reviews the biophysical characteristics of the myocardium compartments included in the proposed numerical phantom. Additionally, in Table (\ref{t:PhantomFeature}), we compare previous numerical phantoms in terms of the biophysical characteristics of the myocardial compartments they incorporated.

\begin{table*}[ht]
\caption{Comparison between the proposed and previous numerical phantoms including biophysical properties of the myocardium}
    \centering
    \begin{tabular}{M{2.7cm}M{1.5cm}M{1.5cm}M{1.5cm}M{1.5cm}M{0.8cm}M{0.8cm}M{0.8cm}M{0.8cm}M{1.2cm}M{.9cm}}
    \hline\hline
    Features & Radial shape variability of CMs & Axial shape variability of CMs & Sarcolemma permeability & ICDs permeability & HA$^\circ$ & SE$^\circ$ & SA$^\circ$ & TA$^\circ$ & Regional curvature & Collagen \\ \hline

    \citep{wang2011simulation} & \XSolidBrush & \XSolidBrush & \Checkmark & \XSolidBrush & \XSolidBrush & \XSolidBrush& \XSolidBrush & \XSolidBrush & \XSolidBrush & \XSolidBrush  \\

    \citep{wang2012multiscale} & \XSolidBrush & \XSolidBrush & \Checkmark & \XSolidBrush  & \Checkmark & \XSolidBrush & \XSolidBrush & \XSolidBrush & \XSolidBrush & \XSolidBrush \\

    \citep{bates2017monte} & \XSolidBrush & \XSolidBrush & \Checkmark  & \XSolidBrush & \Checkmark & \XSolidBrush & \XSolidBrush & \XSolidBrush & \XSolidBrush & \XSolidBrush  \\

    \citep{rose2019novel} & \Checkmark & \XSolidBrush & \XSolidBrush & \XSolidBrush  & \Checkmark & \Checkmark & \XSolidBrush & \XSolidBrush & \XSolidBrush & \XSolidBrush \\
 
    Proposed phantom  & \Checkmark & \Checkmark & \Checkmark & \Checkmark  & \Checkmark & \Checkmark  & \Checkmark & \Checkmark & \Checkmark & \Checkmark\\
    
    \hline\hline
    \end{tabular}
\label{t:PhantomFeature}
\end{table*}

\subsection{CM}
CMs are the cells contributing the most toward the measured dMRI signal in the myocardium.
They occupy 65\textemdash75\% \citep{chen2007method,greiner2018confocal,schaper1985ultrastructural, skepper1995ultrastructural} of the myocardial volume and account for 25\textemdash35\% of all cells \citep{pinto2016revisiting}. 
CMs support the mechanical contraction of the heart, which is necessary to pump blood to all organs within the body \citep{fraticelli1989morphological}. 
\begin{figure}[!tb]
\centering
\includegraphics[width=3.4in]{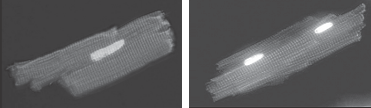}
\caption{Two examples of the shape of an isolated CM with one and two nuclei (Brighter objects) in the left and right, respectively. Images with permission from \citep{gerdes2014assessment}.}
\label{fig:CM}
\end{figure}
\subsubsection{CM Shape} \label{sec:CMdimension} The CM's membrane restricts the movement of water molecules and affects the dMRI signal.
Therefore, we consider a realistic shape of the CMs in our phantom, as they constrain the mobility of water molecules.
As illustrated in Figure \ref{fig:CM}, native CMs do not have a regular and consistent shape. 
However, they are typically approximated as elliptical cylinders \citep{bolli2014manual}.
A standard method to measure the dimensions of CMs is through the combined use of a Coulter channelyser and an optical microscope. 
The average length for the minor-axis ($B$), major-axis ($A$), longitudinal axis ($L$) of CMs along with their volume ($V$) have been reported as $12\pm\SI{1}{\micro\meter}$. $30\pm\SI{1}{\micro\meter}$, $141\pm\SI{9}{\micro\meter}$, and $39933\pm\SI{4640}{\micro\meter^3}$ respectively \citep{chen2007method}.

\subsubsection{CM Membrane Permeability}
The permeability of CMs, defined as the CM-ECS water molecules exchange rate through sarcolemma, or CM-CM water molecules exchange rate through ICDs, is another biophysical feature of CMs that affects the dMRI signal.
\paragraph{Sarcolemma} The sarcolemma is a plasma membrane that acts as a boundary between the ICS cytoplasm and ECS. 
It was shown that, during ischemic injury, the sarcolemma is ruptured, and CM permeability increases \citep{celes2010increased}.
This feature is reflected in the phantom by assigning an interface permeability for the side surface of the CMs. 
\paragraph{ICDs} In adult hearts, ICDs are membranous regions where myofibrils of individual CMs are connected end-to-end with each other.
Gap junctions are ICS channels through the ICDs that allow the passage of small molecules and ions between CMs through ICS diffusion.
This arrangement facilitates the transmission of electrical impulses in the heart. 
Several studies have shown that gap junctions reduce in heart failure and can lead to fatal arrhythmia, leading to a decrease in the permeability of ICDs \citep{pinali2015three, perriard2003dilated, noorman2009cardiac}.
ICDs are modelled by attributing an interface permeability to the top and bottom sides of the CMs.
\subsubsection{CM Arrangement} \label{sec:CMsArrangement} 
\begin{figure}[!tb]
\centering
\includegraphics[width=3.2in]{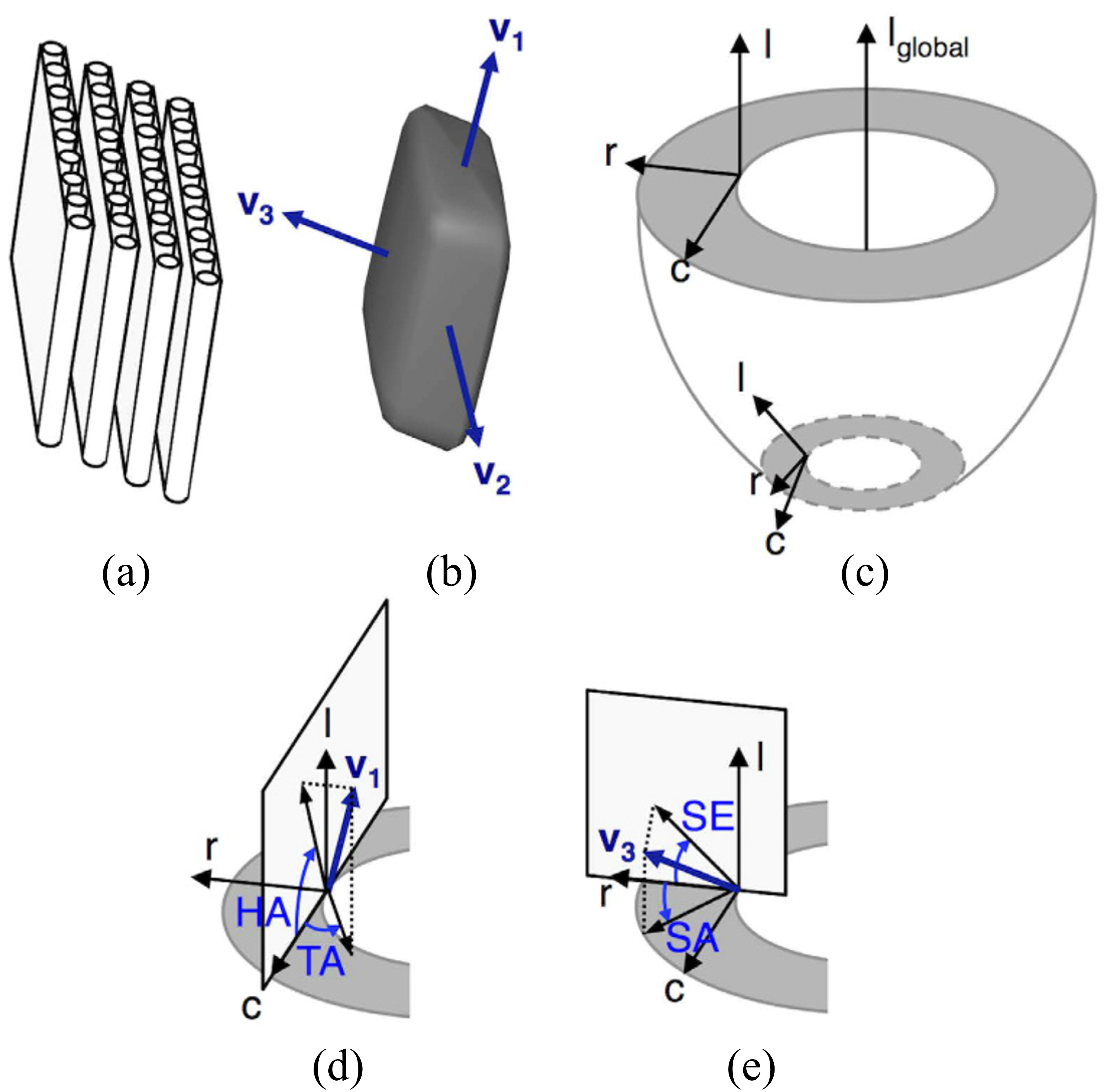}
\caption{Definition of helix, transverse, sheet elevation, and sheet azimuth angles. a) Tightly packing CMs into a laminar structure; b) Superquadric glyph representation of the diffusion tensor and eigenvectors of (a); c) illustration of local cardiac coordinate; d) Helix and transverse angles, e) Sheet elevation and sheet azimuth angles. Image with permission from \citep{teh2016resolving}.}
\label{fig:SheetletAngles}
\end{figure}

The spatial organization of CMs within the myocardium determines biophysical characteristics that affect the dMRI signal.
These include the following:
\paragraph{Sheetlet angles}
Individual CMs are tightly packed into parallel laminar microstructures (as shown in Figure \ref{fig:SheetletAngles}a), approximately 2\textemdash4 CMs thick, referred to as sheetlets \citep{hales2012histo}. 
These sheetlets orient locally in the tissue and are separated by cleavage with a width of 1\textemdash2 CMs from each other \citep{legrice2005architecture}.
It has been shown and validated by histology that the primary (\textbf{$V_1$}), secondary (\textbf{$V_2$}), and tertiary (\textbf{$V_3$}) eigenvectors of cardiac DTI (cDTI) correspond to the CM's long-axis, sheetlet, and sheetlet-normal directions, respectively, as shown in Figure \ref{fig:SheetletAngles}a and \ref{fig:SheetletAngles}b.

\begin{figure}[ht!]
\centering
\includegraphics[width=3.2in]{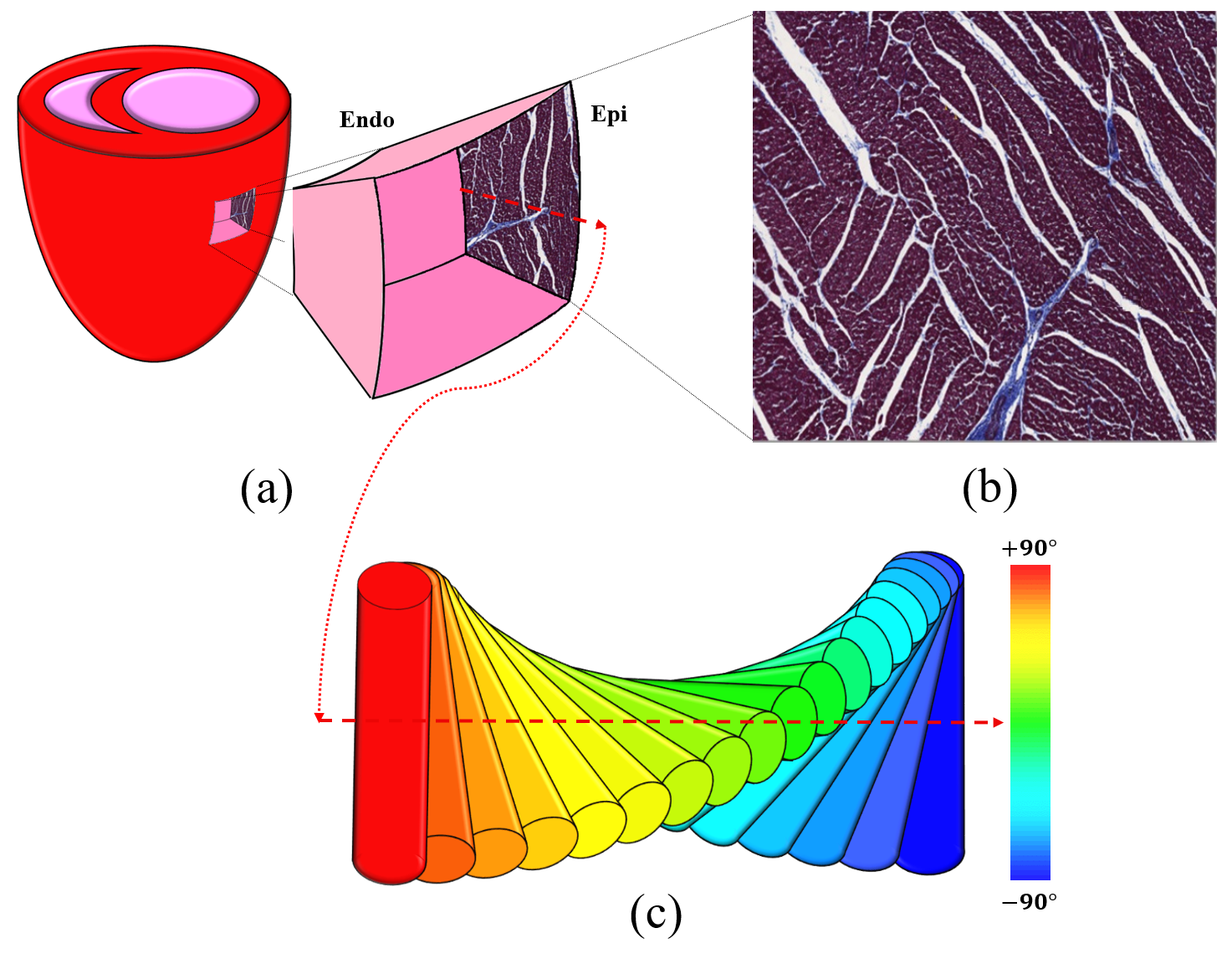}
\caption{The arrangement of CMs in sheetlets in cardiac tissue. (a) a cubic slab of myocardium in local cardiac coordinate; (b) histological image from \textbf{l-r} plane view of myocardium, with permission from \citep{nielles2017assessment}; (c) Simplified schematic of direction changes in CM's long axes transmurally from endocardium to epicardium where the heat-map shows HA variation.}
\label{fig:SheetletAndTwist}
\end{figure}

Using \textbf{$V_1$} and \textbf{$V_3$} \citep{teh2016resolving}, four angles are usually considered to describe these directions of the sheetlets.
As shown in Figure \ref{fig:SheetletAngles}c, a local coordinate system can be defined for each voxel in a cardiac image volume, based on the surface curvature of the cardiac wall, distinguished by longitudinal (\textbf{l}), radial (\textbf{r}) and circumferential (\textbf{c}) axes.
\textbf{r} is computed using the Laplace method \citep{jones2000three}, whereas \textbf{c} is defined as vectors perpendicular to \textbf{r} and global longitudinal ($\textbf{l}_{global}$, a line fitted to the center of left ventricle cavity in 2D short-axis planes).
Finally, \textbf{l} is defined as vectors perpendicular to $\textbf{r}\times\textbf{c}$.
Therefore, as shown in Figure \ref{fig:SheetletAngles}d, HA and \textit{transverse angle} (TA) are defined as the angles subtended by \textbf{c} and projection of \textbf{$V_1$} on \textbf{l-c} and \textbf{r-c} planes, respectively.
The angle subtended by \textbf{r} and projection of \textbf{$V_3$} on \textbf{l-r} and \textbf{r-c} planes is called \textit{sheet elevation} (SE) and \textit{sheet azimuth} (SA) angles.
\paragraph{Myocardial transmural twist} Figure \ref{fig:SheetletAndTwist} represents a simplified schematic of CM's long-axis direction, where it starts with a right-handed helical orientation in the endocardium and smoothly changes toward a left-handed helical orientation in the epicardium \citep{hales2012histo}. 
The measured \textbf{$V_1$} corresponds to an average of the CMs' long axes directions within an MRI voxel \citep{streeter1969fibre}.
Therefore, the magnitude of the primary eigenvalue of cDTI should be modulated by the standard deviation of this helical orientation distribution.

\paragraph{Ventricle curvature} Ventricle curvature is another important geometrical feature that affects sheetlets' shape, and consequently, CMs' shape (of myocardial tissue that undergoes remodelling in some CVDs). 
For instance, a normal left ventricle has an ellipsoidal geometry that alters to a more spherical one following myocardial infarction \citep{su2012geometrical}. 
The curvature values for different regions of the healthy and infarcted myocardium of rats are reported in \citep{weisman1985global,espe2017regional}.

\subsection{Collagen}
\begin{figure}[ht!]
\begin{center}
    \begin{subfigure}[normal]{.49\linewidth}
    \centering
    \includegraphics[width=1\linewidth]{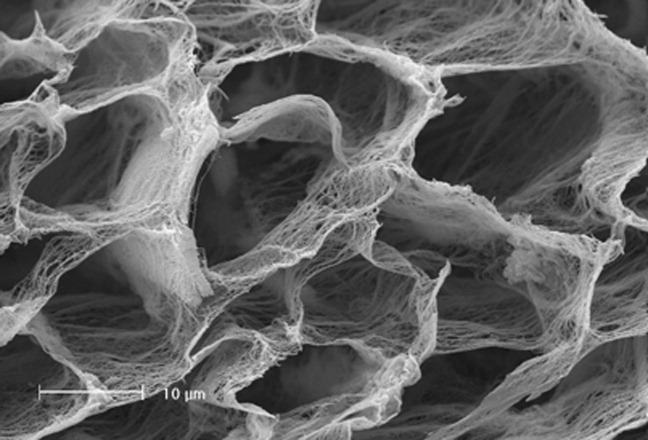}
    \caption{}
    \label{fig:Endomysial}
    \end{subfigure}
    \begin{subfigure}[normal]{.49\linewidth}
    \centering
    \includegraphics[width=1\linewidth]{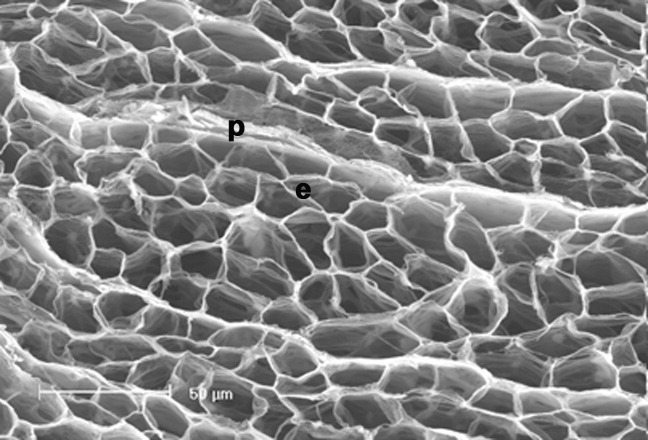}
    \caption{}
    \label{fig:EndoPeri-mysial}
    \end{subfigure}
\end{center}
\caption{SEM of the three-dimensional arrangement of the cardiac collagen fibres in healthy dogs. a) Endomysia b) Endomysia (e) and Perimysium (p). Image with permission \citep{benedicto2011structural}.}
\label{fig:collagen}
\end{figure}

Collagen fibre is the dominant component of ECM. 
It is responsible for providing structural support by transmitting forces, preventing overstretching and rupture, preserving the shape and thickness of the myocardium, and providing both active and passive stability to the myocardium \citep{benedicto2011structural}.
These fibres fall into three categories: endomysium, surrounding and interconnecting individual CMs and capillaries; perimysium, surrounding and interconnecting groups of myocytes; and epimysium, surrounding the entire muscle \citep{pope2008three}.
Of the collagen fibres, only endomysium (Figure \ref{fig:collagen}a) and perimysium (Figure \ref{fig:collagen}b) are modelled in the proposed phantom as it is unlikely for many dMRI voxels to include effects of the epimysium in the dMRI signal. 
Furthermore, since the thickness of the collagen fibre is at a nanoscale \citep{benedicto2011structural}, given the typical diffusion time scale of MRI and the consequent diffusion of water molecules, the effect of collagen fibre on ECS diffusion is coarse-grained. 
Therefore, the homogenising effect of the collagen fibre is seen as an effective medium with specific diffusivity \citep{novikov2016quantifying}.

\section{Design of Numerical Phantom} \label{sec:Design}
As discussed earlier, the arrangement of CMs in the myocardium can be considered at two scales, i.e., at the scale of a sheetlet and at the scale of the myocardial wall (comprising several sheetlets). 
At the sheetlet scale, CMs are densely packed with a near-parallel arrangement into a sheet-like structure \citep{hales2012histo}. 
Conversely, the sheetlets are placed next to each other at the wall scale while orientated based on desired sheetlet angles.
Therefore, packing CMs with diverse shapes while preserving the input PDFs of the shape parameters is a challenging task. 
This section proposes a method to address this task in two stages, corresponding to the sheetlet and wall scales.
\begin{figure*}[tb]{}
\centering
\includegraphics[width=7in]{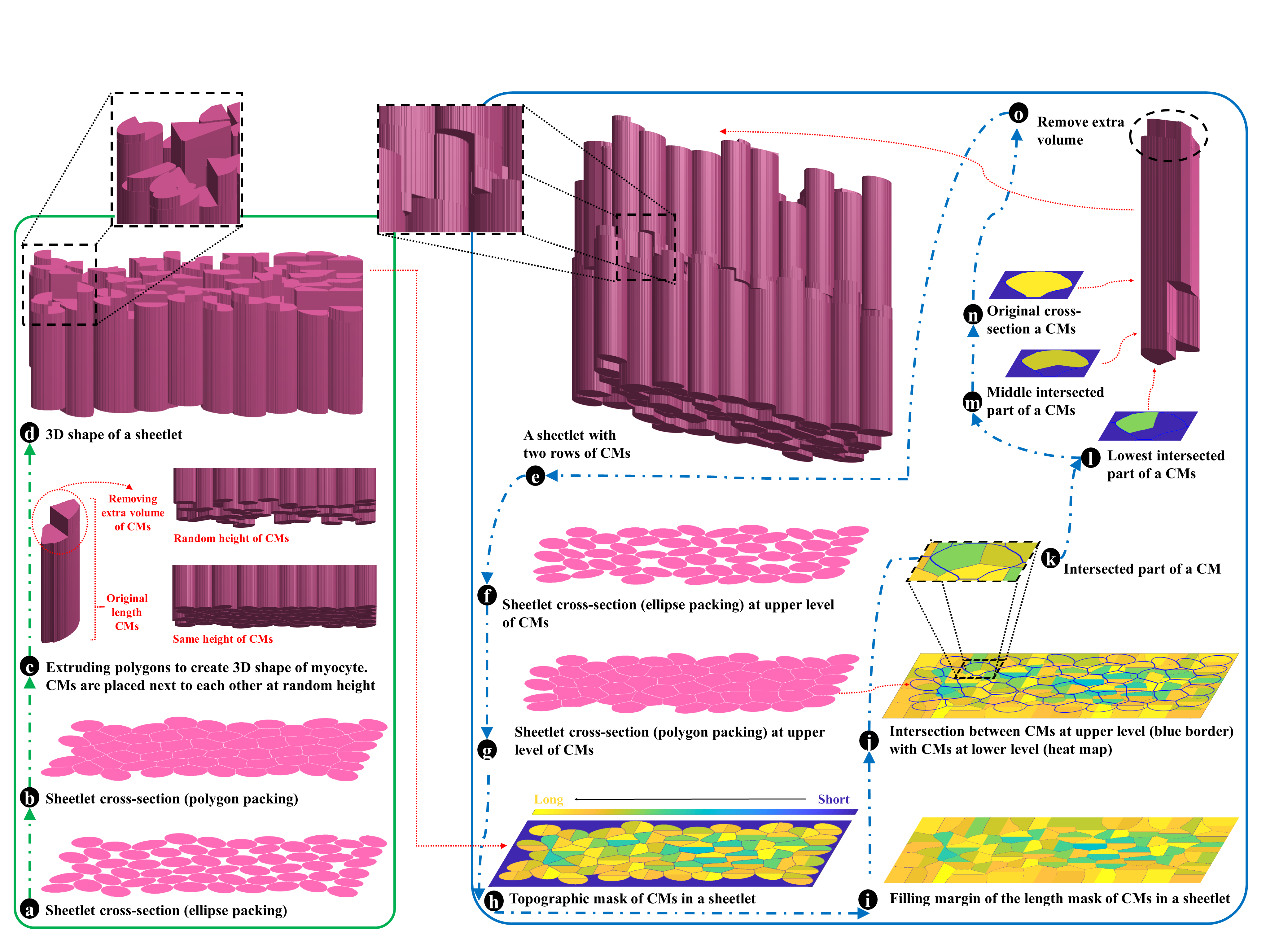}
\caption{Schematic overview of the proposed approach for generating a numerical phantom of the myocardium, at the sheetlet scale.}
\label{fig:SheetletAlgorithm}
\end{figure*}

\subsection{Sheetlet Scale}
\subsubsection{Generating a 2D Cross-Section of a Sheetlet}
    We assume a fully parallel arrangement of the CMs in the sheetlets, based on empirical evidence \citep{hales2012histo}. 
    Therefore, dense packing of the CMs in a sheetlet can be effected in 2D and subsequently in 3D:
        \renewcommand{\labelenumi}{\roman{enumi}}
        \begin{enumerate}
        \item\textit{2D Ellipse Packing:} 
        Limitation of the CMs' simplified cross-sections to ellipses or circles render a dense ellipse packing algorithm well suited to our 2D problem. 
        Here we use the dense ellipse packing algorithm proposed in \citep{ilin2016advancing} to generate a primary 2D cross-section of the CMs packed densely in a 2D plane (Figure \ref{fig:SheetletAlgorithm}a).
        The packing domain of ellipses is rectangular with a width of sheetlet thickness (Table \ref{t:PhantomFeature}) and a length, larger than the final voxel's length.
        In this study, we set the length of the packing domain to 200 $\SI{}{\micro\meter}$, two times larger than the final voxel's length ($\SI{100}{\micro\meter}$).
        The inputs to this algorithm are $V$, $L$, $A$, $B$, and dimensions of the packing domain (sheetlet thickness and length), shown in Table \ref{t:PhantomFeature}. 
        \item\textit{Transformation of the 2D Pack of Ellipses to Polygons:} According to histological observations \citep{bensley2016three}, the native cross-section of CMs are polygons. 
        Therefore, as proposed in \citep{st20083d}, the pack of ellipses is transformed into polygons using a watershed algorithm. 
        Figure \ref{fig:SheetletAlgorithm}.b depicts the cross-section of a myocardial sheetlet obtained from transforming packed ellipses to polygons.
        \end{enumerate}

\subsubsection{Generating the 3D Shape of a Sheetlet Based on the 2D Cross-Section} \label{sec:removesmallsegment}
    The 3D shape of a sheetlet is formed by extruding the border of polygons/CMs along the normal axis to the cross-section of the sheetlet, according to the length PDF of CMs.
    Preserving the length PDF of CMs changes the volume PDF of CMs due to an increase in the area of the CMs' cross-section, following the transformation of ellipses to polygons. 
    Hence, to preserve both input length and volume PDFs of CMs, a small segment is removed from the top of the extruded CMs as depicted in Figure \ref{fig:SheetletAlgorithm}c (left).
    This small segment is made up of the half cross-section of the CM, extruded long enough to make its volume equal to the extra volume of the CM.
    Moreover, to better emulate real tissue and maintain some irregularity in the arrangement of CMs within the phantom, CMs are placed at random heights (Figure \ref{fig:SheetletAlgorithm}c, right).
    For each CM, the random height is a positive number, chosen from a uniform distribution in which the upper end of the CM does exceed the height of the highest CM in each group of CMs.
    The 3D shape of a short section of a sheetlet is illustrated in Figure \ref{fig:SheetletAlgorithm}d.
    
\subsubsection{Generating a Sheetlet} 
Given the average length of the CMs, $141\pm9$ $\mu m$, and the size of the in-silico voxel, several layers of the CMs must be stacked on top of one another to ensure a complete sheetlet is long enough.
Figure \ref{fig:SheetletAlgorithm}e shows a short section of a sheetlet comprising two layers of CMs.
The main challenges for extending a sheetlet, layer by layer, is that the ends of CMs at the lower layer (LL) are not aligned, as depicted in Figure \ref{fig:SheetletAlgorithm}d (magnified).
To model ICDs and keep the ECS more realistic, each CM of the upper layer (UL) at its bottom should have a complementary shape concerning the top of the CMs in the LL, as shown in Figure \ref{fig:SheetletAlgorithm}e (magnified).
This part of the algorithm includes the following steps:
        \begin{enumerate}
        \item\textit{Designing a Complementary Shape for CMs of UL:}
        First, to complete the sheetlet, along with preserving the randomness of the distribution of the CMs in the phantom, for each layer, a new cross-section of CMs are generated as described in Figure \ref{fig:SheetletAlgorithm}f and \ref{fig:SheetletAlgorithm}g. 
        Second, information about the intersections between the CMs in UL with CMs in LL and the height of their intersections, are acquired. 
        Subsequently, using the acquired information, the complementary shape of CMs in UL is constructed.
        Next, we describe the required steps in detail:
            \begin{enumerate}
            \item\textit{Finding the Intersection Between the CMs in the Successive Layers:} 
            First, a topographic map of CMs in LL is generated as depicted in Figure \ref{fig:SheetletAlgorithm}h.
            Next, since the cross-section of CMs in LL and UL are not the same (Figure \ref{fig:SheetletAlgorithm}b vs. \ref{fig:SheetletAlgorithm}g), it is likely that some parts of marginal CMs in the UL, Figure \ref{fig:SheetletAlgorithm}g, intersect the blue region (vacant region) in Figure \ref{fig:SheetletAlgorithm}h. 
            Therefore, it is impossible to find the starting height for those parts of marginal CMs in UL, which intersect the blue region in Figure \ref{fig:SheetletAlgorithm}h.
            Hence, the topographic map is modified so that the blue region is filled by the extension of marginal CMs in LL, as shown in Figure \ref{fig:SheetletAlgorithm}i.
            Then, each CM in UL is intersected with the modified topographic map. 
            The result of the intersection of CMs in UL with the topographic map of CMs in LL is as shown in Figure \ref{fig:SheetletAlgorithm}j, where blue polygons indicate the border of CMs of UL, shown in Figure \ref{fig:SheetletAlgorithm}g.
            
            \item\textit{Forming 3D Shape of CMs in UL:}
            For the intersection of each CM in UL with CMs in LL (e.g., Figure \ref{fig:SheetletAlgorithm}k, a blue polygon with topographic map), the lowest segment (Figure \ref{fig:SheetletAlgorithm}l, the intersection of green region with blue polygon) is extruded to the height of second-lowest segment (Figure \ref{fig:SheetletAlgorithm}m).
            Then, at the height of the second-lowest segment, the first- and second-lowest segments (green and dark yellow inside the blue border, in Figure \ref{fig:SheetletAlgorithm}k) are merged, as illustrated in Figure \ref{fig:SheetletAlgorithm}m, and extruded to the height of the third-lowest segment (Figure \ref{fig:SheetletAlgorithm}n).
            This procedure is repeated until all segments of a CM are merged, and the original cross-section of the CM is retrieved (Figure \ref{fig:SheetletAlgorithm}n). 
            Next, the original cross-section of CM is extruded long enough to preserve the input length of the CM.
            Finally, for CMs with increased volumes, a small segment is removed, illustrated in Figure \ref{fig:SheetletAlgorithm}o, as explained in \ref{sec:removesmallsegment}.  
            \end{enumerate}
        \end{enumerate} 
        
Figure \ref{fig:SheetletAlgorithm}e shows adding the second layer of the CMs to the first layer (Figure \ref{fig:SheetletAlgorithm}d).
A complete sheetlet is formed by repeating this algorithm, blue box in Figure \ref{fig:SheetletAlgorithm}, for a desired number of layers.

\subsection{Wall Scale}
\begin{figure*}[tb]
\centering
\includegraphics[width=6in]{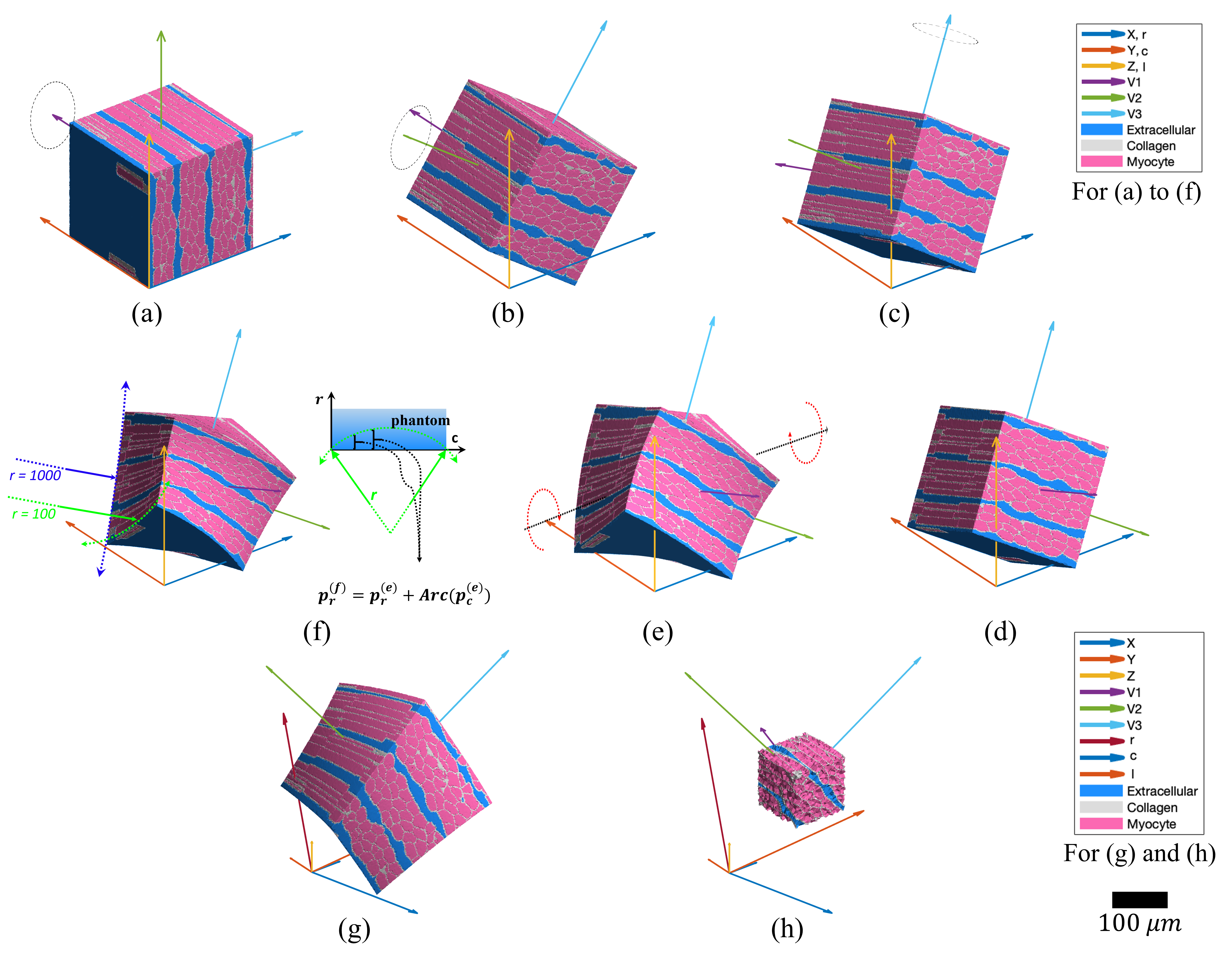}
\caption{Algorithm for generating myocardium numerical phantom at wall scale: a) initial status of the sheetlets in-wall configuration, where \textbf{$V_1$}, \textbf{$V_2$}, \textbf{$V_3$} are parallel to \textbf{c}, \textbf{l}, \textbf{r}, respectively; b) modelling SE$^\circ$ by rotating phantom in (a) around \textbf{$V_1$}; c) modelling SA$^\circ$ by rotating phantom in (b) around \textbf{$V_2$}; d) modelling HA$^\circ$ by rotating phantom in (c) around \textbf{$V_3$}; e) Twisting the phantom the axis crossing the centre of the phantom and parallel to \textbf{r}; f) modelling wall curvature by bending the phantom in (e) around the axes \textbf{l} and \textbf{c}; g) Transforming the phantom in (f) to cardiac local coordinate to mimic the sheetlet orientations in ex-vivo data; h) Extracting a cubic voxel to mimic an MRI voxel.}
\label{fig:WallAlgorithm}
\end{figure*}

This Section describes the process of generating a voxel of the myocardium at a microscale, using the generated sheetlets in the previous Section.
First, several sheetlets, depicted in pink in Figure \ref{fig:WallAlgorithm}a, are placed next to each other in the ECS, revealed by the blue color in Figure \ref{fig:WallAlgorithm}a.
Endomysial and perimysium collagen fibres domains are indicated by grey colour in Figure \ref{fig:WallAlgorithm}a.
The boundary of the collagen fibres domain is defined as an enlarged outline of the sheetlet.
The ECS inside this domain, except the ICS regions, is labelled as the collagen region, and its effect is taken into account by assigning diffusivity ($D_{collagen}$) and relaxation ($T_{2_{collagen}}$) of collagen to it.

Next, the sheetlet angles are modelled by aligning the phantom in the directions of the desired eigenvectors.
Because the eigenvectors are orthogonal, one only needs to align two of them in the desired directions.
Since the SE, SA, HA, and TA angles are defined for \textbf{$V_1$} and \textbf{$V_3$}, we choose these eigenvectors for the alignment.
Again, based on the orthogonality of eigenvectors, by setting three sheetlet angles out of four, \textbf{$V_1$} and \textbf{$V_3$} and consequently, \textbf{$V_2$} are aligned in the desired directions.
Here, we orient the phantom based on SE, SA, and HA angles.
\par Initially, according to the relationship between sheetlet orientations and eigenvectors of cDTI, as shown in Figure \ref{fig:SheetletAngles}b, \textbf{$V_1$}, \textbf{$V_2$} and \textbf{$V_3$} are set as depicted in Figure \ref{fig:WallAlgorithm}a.
Then, to align \textbf{$V_3$} in the desired direction, first, the phantom in its initial status, Figure \ref{fig:WallAlgorithm}a, is rotated around \textbf{$V_1$} so that the projection of \textbf{$V_3$} on \textbf{l-r} makes SE$^\circ$ angle with \textbf{r}, shown in Figure \ref{fig:SheetletAlgorithm}b.
Second, the phantom is rotated around \textbf{$V_2$}, Figure \ref{fig:WallAlgorithm}b, so that SA$^\circ$ is subtended by \textbf{r} and the projection of \textbf{$V_3$} on \textbf{r-c} as shown in Figure \ref{fig:WallAlgorithm}c.
Then, to align \textbf{$V_1$} in the desired direction, we only need to rotate the phantom around \textbf{$V_3$}, Figure \ref{fig:WallAlgorithm}c. As a result, the subtended angle by the projection of \textbf{$V_1$} on \textbf{l-c} and \textbf{c}, illustrated in Figure \ref{fig:WallAlgorithm}d, is equal to the desired HA$^\circ$.
\par Next, the tissue is twisted by rotating the phantom of Figure \ref{fig:WallAlgorithm}d around the axis parallel to \textbf{r}, crossing through the center of the phantom, shown in Figure \ref{fig:WallAlgorithm}e, in which the degree of rotation ($\alpha$) is increased by increasing the distance from the center of the phantom, as depicted in Figure \ref{fig:WallAlgorithm}e. 
\par Then, the curvature of the myocardium wall is included in the phantom.
For this purpose, the phantom should be bent around an axis parallel to \textbf{c} and crossing the point of $(\textbf{c}, \frac{l}{2}, 0)$, and an axis parallel to \textbf{l} and crossing through $(\frac{l}{2}, \textbf{l}, 0)$, where ($l$) is the length of the phantom across the axis \textbf{c}.
For example, for the latter, all points in every \textbf{c-r} plane, depicted as blue in Figure \ref{fig:WallAlgorithm}f on the right-hand side, should be uniformly relocated to the red-dashed arrow. 
As depicted in Figure \ref{fig:WallAlgorithm}f (right-hand side), using the values of curvature $K=\frac{1}{r}$, where $r$ (the green arrow) is the radius of the curvature, the displacement along axis \textbf{r}, $d_{r}=Arc(p_{c})$, is computed where $Arc$ computes the displacement of the location of phantom's nodes, i.e., $(p_{r},p_{c},p_{l})$, along the axis \textbf{r} using $p_{c}$ or $p_{l}$.
Then, the phantom is bent around an axis parallel to \textbf{l} and crossing through $(\frac{l}{2}, \textbf{l}, 0)$ by updating the location of the phantom's nodes along with \textbf{r} concerning \textbf{c}, i.e., $p_{r}^{(f)}=p_{r}^{(e)} + Arc(p_{c}^{(e)})$ where superscripts $(e)$ and $(f)$ indicate nodes' location in the related phantoms in Figure \ref{fig:WallAlgorithm}e and \ref{fig:WallAlgorithm}f (left-hand side), respectively.
Figure \ref{fig:WallAlgorithm}f (left-hand side) illustrates the bent phantom around an axis parallel to \textbf{l} and crossing from the point of $(\frac{l}{2}, \textbf{l}, 0)$ indicated by a light green arc with $r=100$.
Similarly, by updating $p_{r}^{(f)}=p_{r}^{(e)} + Arc(p_{l}^{(e)})$, the phantom is bended around the axis parallel to \textbf{c} and crossing the point of $(\textbf{c}, \frac{l}{2}, 0)$, shown in Figure \ref{fig:WallAlgorithm}f (left-hand side) as blue arc with $r=1000$.
\par In the next step, only required for mimicking an MRI voxel, the phantom is transformed into a local coordinate system (Figure \ref{fig:WallAlgorithm}g).
Finally, as shown in Figure \ref{fig:WallAlgorithm}h, a cubic slab which its sides are parallel to the axes of the global coordinate system is extracted from Figure \ref{fig:WallAlgorithm}g to represent an MRI voxel.
\par
\begin{figure}[ht!]
\begin{center}
    \begin{subfigure}[normal]{.37\linewidth}
    \centering
    \includegraphics[width=1\linewidth]{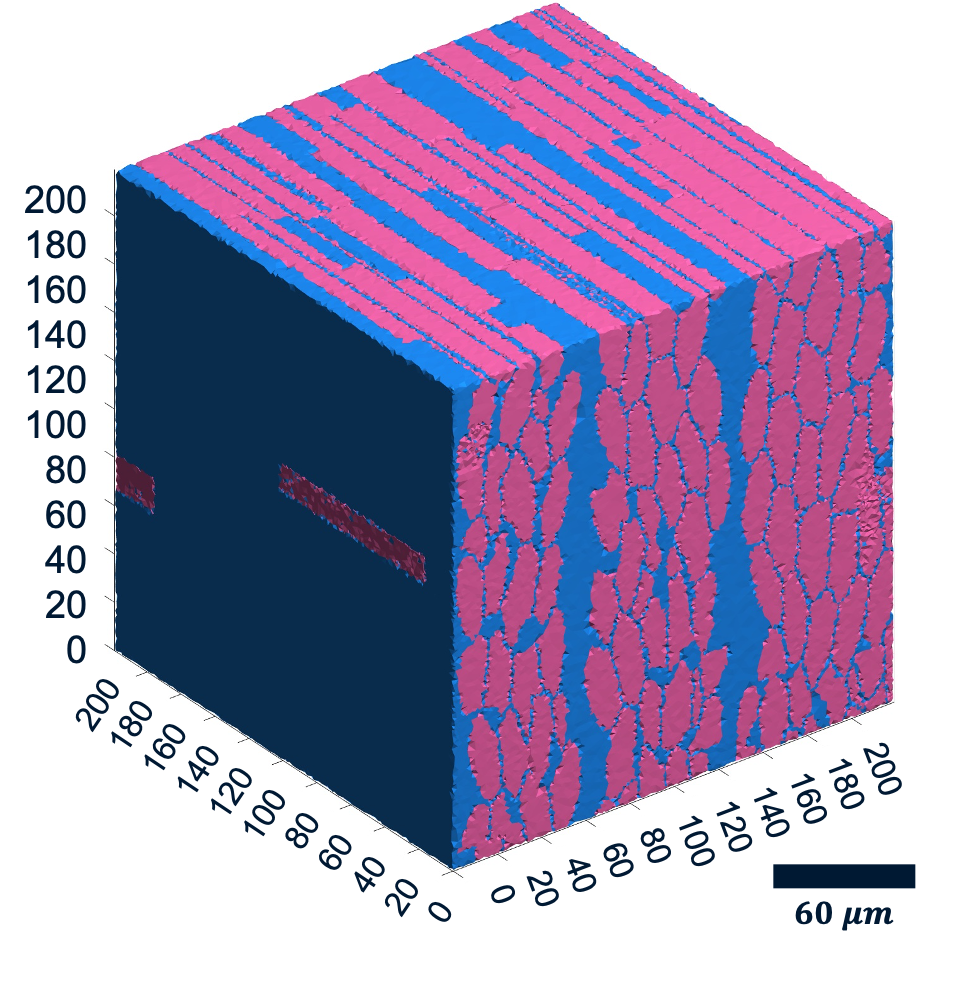}
    \caption{}
    \label{fig:0InterSheet1.0InterCM}
    \end{subfigure}
    \begin{subfigure}[normal]{.37\linewidth}
    \centering
    \includegraphics[width=1\linewidth]{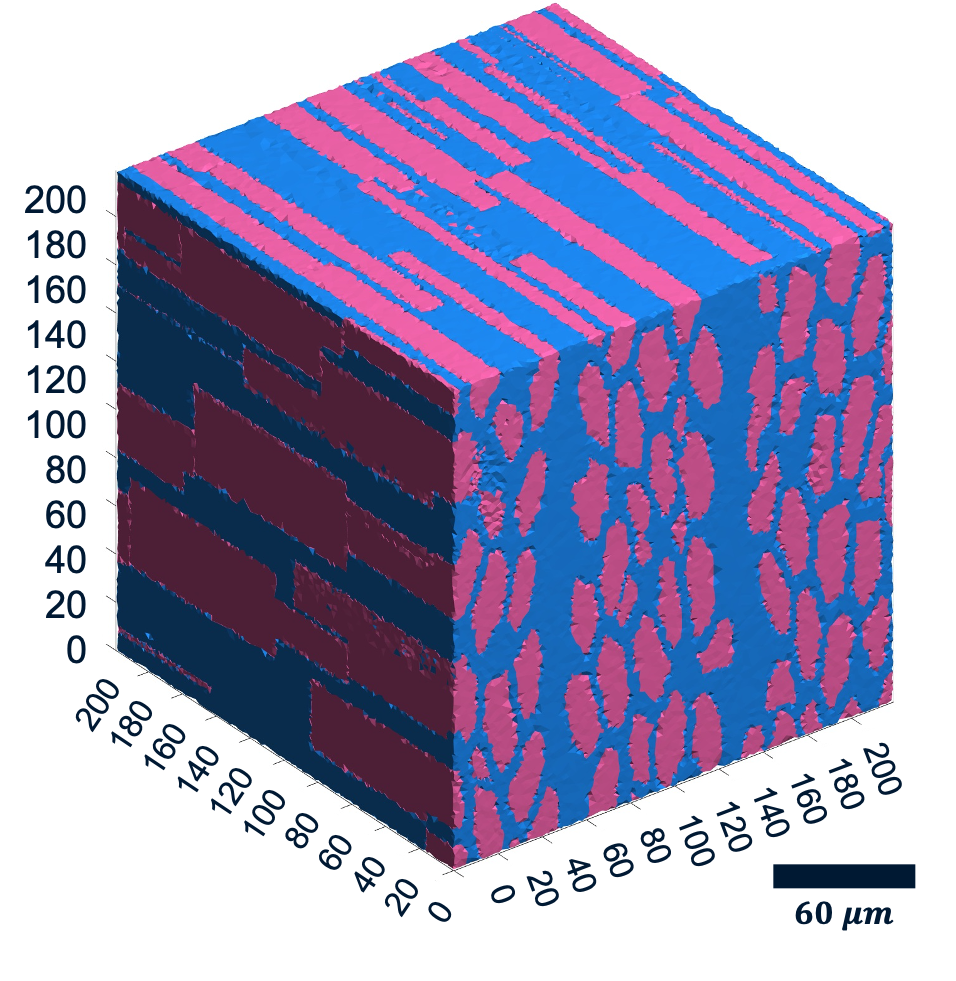}
    \caption{}
    \label{fig:0InterSheet1.2InterCM}
    \end{subfigure}
    \begin{subfigure}[normal]{.37\linewidth}
    \centering
    \includegraphics[width=1\linewidth]{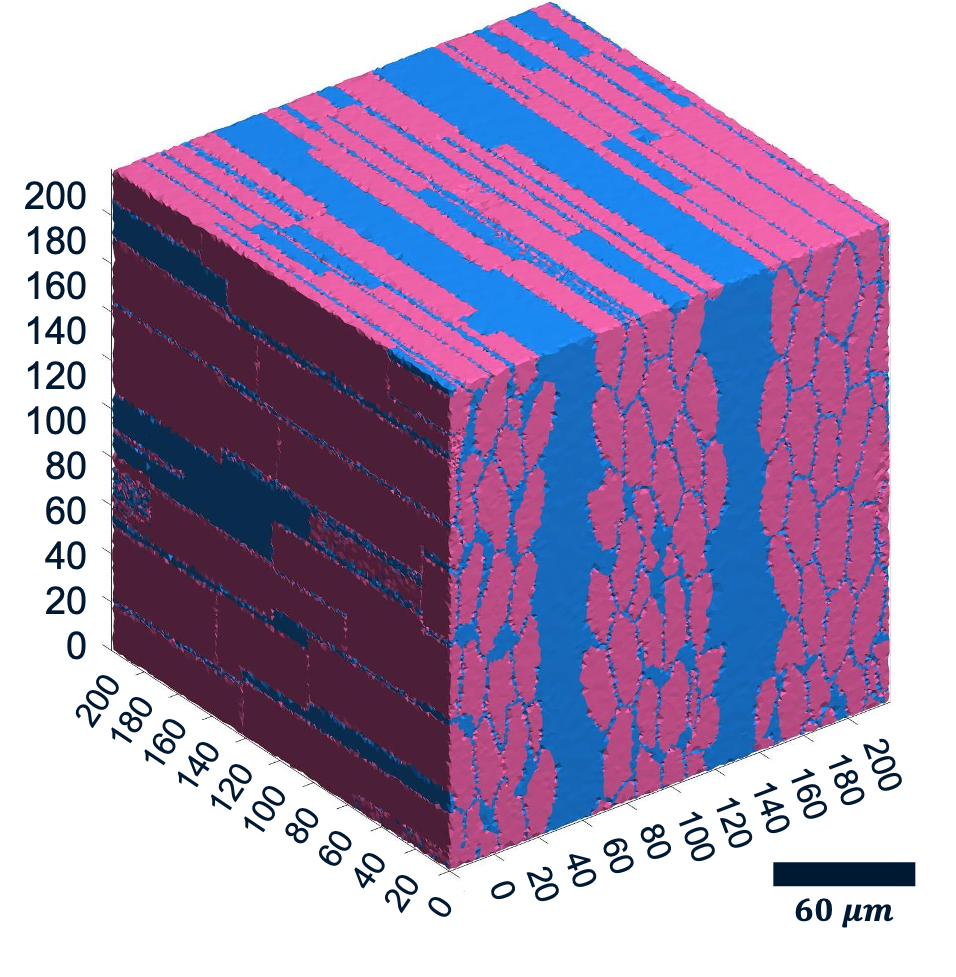}
    \caption{}
    \label{fig:10InterSheet1.0InterCM}
    \end{subfigure}
    \begin{subfigure}[normal]{.37\linewidth}
    \centering
    \includegraphics[width=1\linewidth]{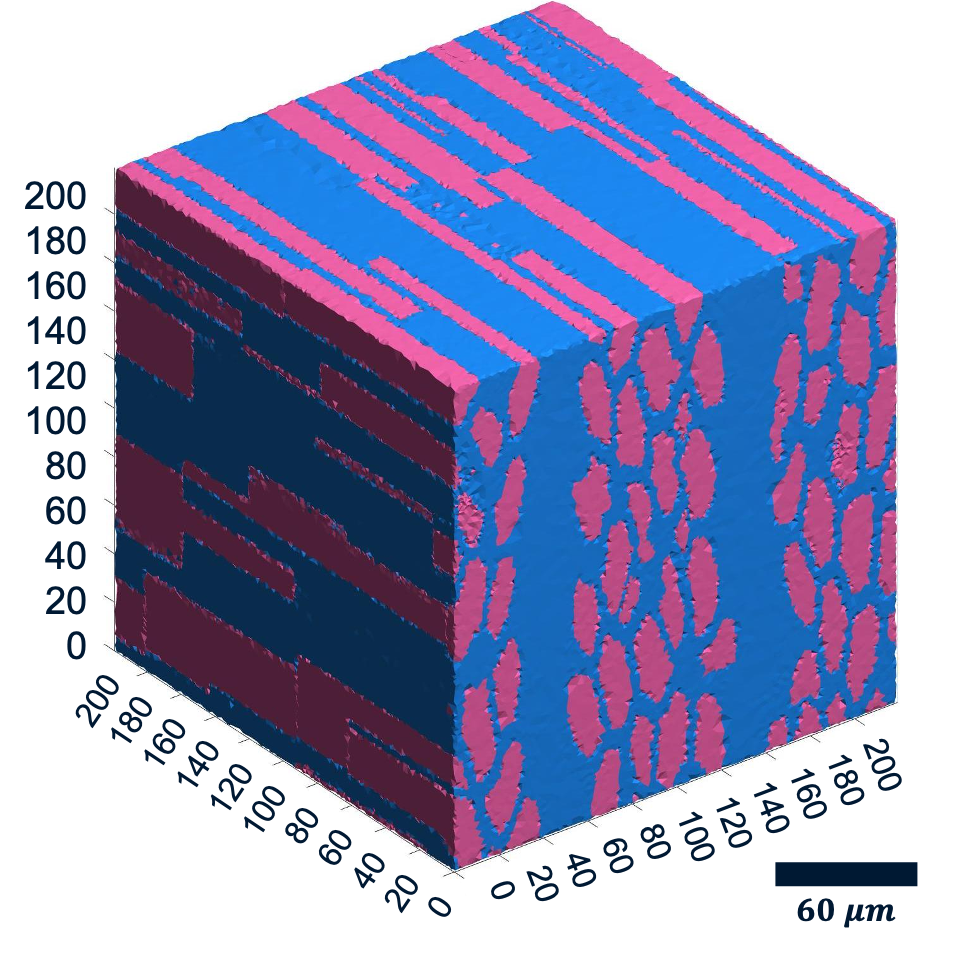}
    \caption{}
    \label{fig:10InterSheet1.2InterCM}
    \end{subfigure}
\end{center}
\caption{Increase of ECS: a) Initial status, b) Increase of inter-CMs space, c) Increase of inter-sheetlet space, d) Increase of both inter-CMs and inter-sheetlet spaces.}
\label{fig:increase of ECS}
\end{figure}

 ECS can be increased by increasing inter-CMs and/or inter-sheetlet spaces.
To increase inter-CMs space, the space between polygons generated in step Figure \ref{fig:SheetletAlgorithm}.b, needs to increase.
The result of increasing inter-CMs space in the phantom is visible by comparing Figure \ref{fig:0InterSheet1.0InterCM} and \ref{fig:0InterSheet1.2InterCM}.
The result of increasing inter-sheetlet space by placing generated sheetlets further apart from each other is distinguishable by comparing Figure \ref{fig:0InterSheet1.0InterCM} vs. \ref{fig:10InterSheet1.0InterCM}.
Finally, the result of increasing inter-CMs space along with inter-sheetlet space is shown in Figure \ref{fig:10InterSheet1.2InterCM}.

\section{Experimental Data and Software}
\subsection{Data}\label{sec:data}
The phantom was verified using several voxels of the 3D data acquired in ex-vivo rat hearts reported previously \citep{teh2016resolving}.
Briefly, dMRI data was acquired on a \SI{9.4}{T} preclinical MRI scanner (Agilent, CA, USA) using a 3D fast spin-echo DW sequence with the following parameters: TR/TE = \SI[parse-numbers=false]{250 / 9.3}{ms}, echo train length = 8, echo spacing = \SI{4.9}{ms}, field-of-view = $20\times16\times$\SI{16}{\milli\meter}, resolution = $100\times100\times$\SI{100}{\micro\meter}, diffusion duration ($\delta$) = \SI{2}{ms}, diffusion time ($\Delta$) = \SI{5.5}{ms}, number of non-DW images = 8, number of DW directions = 61, and b-value (effective) =  \SI{1000}{s\per\milli\meter^2}.
The abovementioned values are the same for all the simulations in this work.
\subsection{Software}
All processes including the generation of the numerical phantom, meshing, simulation of dMRI signal, calculation of cDTI parameters, and statistical analysis were carried out in MATLAB (MathWorks, Massachusetts, USA).
\subsubsection{iso2mesh}
iso2mesh is an open-source MATLAB mesh generation and editing toolbox.
It creates a 3D tetrahedral finite element mesh from 3D gray-scale volumetric image, resulting from Section \ref{sec:Design} \citep{fang2009tetrahedral,tran2020improving}.
The different compartments of the numerical phantom i.e., CMs' groups, ECS, and collagens are labelled differently in the resultant mesh.
These labels enable the FEM-based dMRI simulator to resolve diffusivity and relaxation assigned to each compartment, along with permeability between them for the simulation.

\subsubsection{SpinDoctor}
\par SpinDoctor is a free FEM-based MATLAB toolbox that solves the BT\textemdash PDE to simulate the dMRI signal.
BT\textemdash PDE describes the evolution of the transverse magnetization signal mathematically.
It relates the temporal evolution of the transverse magnetization ($M(\textit{\textbf{r}},t)$) to the spatial derivatives, the diffusion coefficient ($D(\textit{\textbf{r}})$), the $T_{2l}$ spin-spin relaxation, and the time-varying magnetic field gradient ($f(t)\textit{\textbf{g}}\cdot\textit{\textbf{r}}$), where $f(t)$ is the effective time profile, and \textit{\textbf{g}} defines the amplitude and direction of the magnetic field gradient. 
Let $\Omega$ be the observation domain, comprising $L$ sub-domains, such that $\cup^{L}_{l=1}\Omega_l$. 
Also, let $\Gamma^{e}_{l}$ be the external boundary of $\Omega_l$, and $\Gamma_{ln}$ the boundary between $\Omega_{l}$ and $\Omega_{n}$. Then, the evolution of the complex transverse magnetisation in the rotating frame is described by
    \begin{equation}\label{eq:BT}
        \begin{split}
            \frac{\partial}{\partial t}M_l(\textit{\textbf{r}},t)&= \nabla\cdot(\textit{D}_l(\textit{\textbf{r}})\nabla M_l(\textit{\textbf{r}},t))
            -\frac{1}{T_{2l}}M_l(\textit{\textbf{r}},t) \\
            & -i\gamma f(t)\textit{\textbf{g}}\cdot\textit{\textbf{r}}M_l(\textit{\textbf{r}},t) \\
            &(\textit{\textbf{r}}\in \Omega_l),
        \end{split}
    \end{equation}
subject to the boundary conditions
    \begin{equation}\label{eq:BC1}
        \begin{split}
            &D_l(\textit{\textbf{r}})\nabla M_l(\textit{\textbf{r}},t)).\textit{\textbf{n}}_l(\textit{\textbf{r}}) = k_{ln}(M_n(\textit{\textbf{r}},t)-M_l(\textit{\textbf{r}},t))\\
            &(\textit{\textbf{r}}\in\Gamma_{nl},\forall n),
        \end{split}
    \end{equation}
    \begin{equation}\label{eq:BC2}
        \begin{split}
            &D_l(\textit{\textbf{r}})\nabla M_l(\textit{\textbf{r}},t)).\textit{\textbf{n}}_l(\textit{\textbf{r}}) = -k^{e}_{l}M_l(\textit{\textbf{r}},t)\\
            &(\textit{\textbf{r}}\in\Gamma^{e}_{l}),
        \end{split}
    \end{equation}
and the initial condition
    \begin{equation}\label{eq:IC}
        \begin{split}
            &M_l(\textit{\textbf{r}},0)) = p_l(\textit{\textbf{r}}),
        \end{split}
    \end{equation}
where $t \in [0,T_E]$, with $T_E$ the echo time, $\gamma$ is the gyromagnetic ratio of protons ($2.675\times10^8$ rad $T^{-1}s^{-1}$ for $^{1}$H), $\textit{\textbf{n}}_l(\textit{\textbf{r}})$ is the unitary outward pointing normal to $\Omega_l$, $k_{ln}$ ($k^e_l$) is the permeability constant in $\Gamma_{ln}$ ($\Gamma^e_{l}$). Also, the same permeability is assumed in both directions of the same membrane, i.e $k_{ln} = k_{nl}$.
Here, to simulate the dMRI signal, we use the SpinDoctor simulator \cite{li2019spindoctor}, which solves the Bloch-Torrey PDEs using the FEM.
\par For all simulations, the ordinary differential equation (ODE) is solved using theta time stepping method (generalized midpoint) \citep{stuart1991dynamics}, with the following parameter setup of SpinDoctor toolbox:
\begin{itemize}
    \item \textit{implicitness} = $\SI{0.5}{}$: calls Crank Nicolson method \citep{stuart1991dynamics};
    \item \textit{timestep} = $\SI{5}{\micro\second}$: timestep for iterations.
\end{itemize}


\section{Experiments and Results}
The quality of the proposed numerical phantom was evaluated in the following experiments: 
(a) qualitative/visual comparison of in-silico CMs and myocardial tissue with their real counterparts;
(b) comparison of the shape of in-silico CMs with real CMs, using a virtual morphometric study;
(c) comparison of microstructure complexity of the in-silico tissue against histology;
and d) verification of cDTI parameters of the simulated cDTI with their experimental counterparts from ex-vivo measurements.
 
\subsection{Qualitative Comparison of In-silico CMs with Real Myocardium Experimental Data}
\begin{figure}[!tb]
     \centering
     \begin{subfigure}[b]{0.4\textwidth}
         \centering
         \includegraphics[width=1\linewidth]{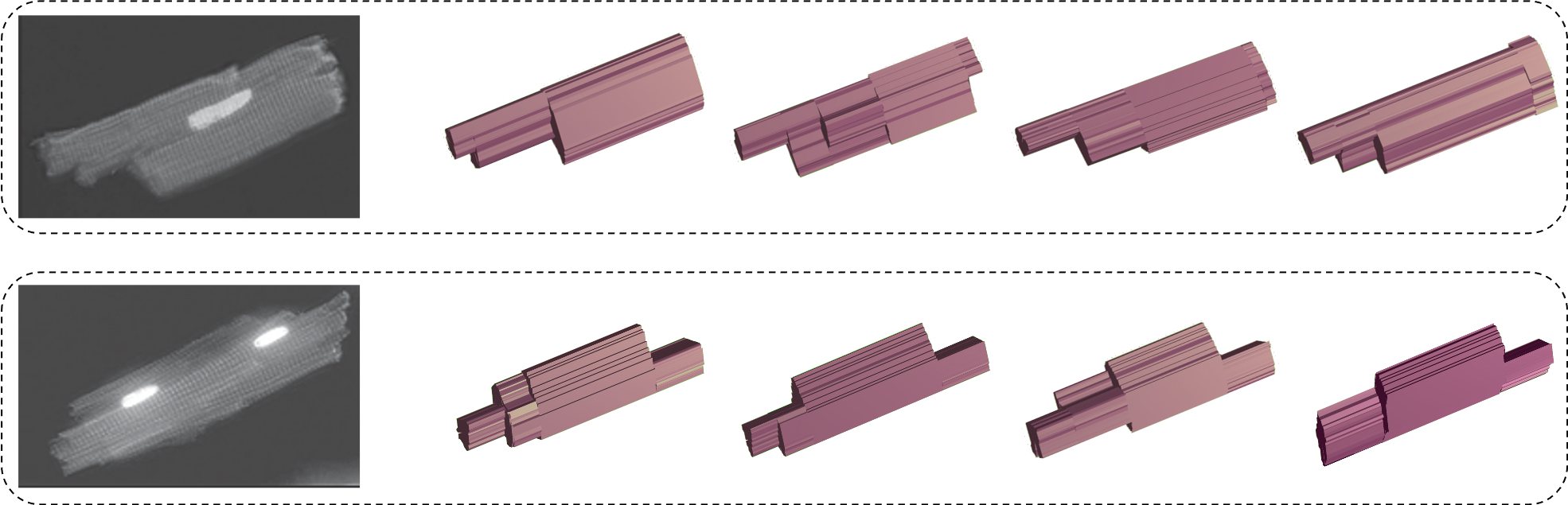}
         \caption{}
         \label{fig:CMcomparison}
     \end{subfigure}
     \hfill     
     \begin{subfigure}[b]{0.4\textwidth}
         \centering
         \includegraphics[width=1\linewidth]{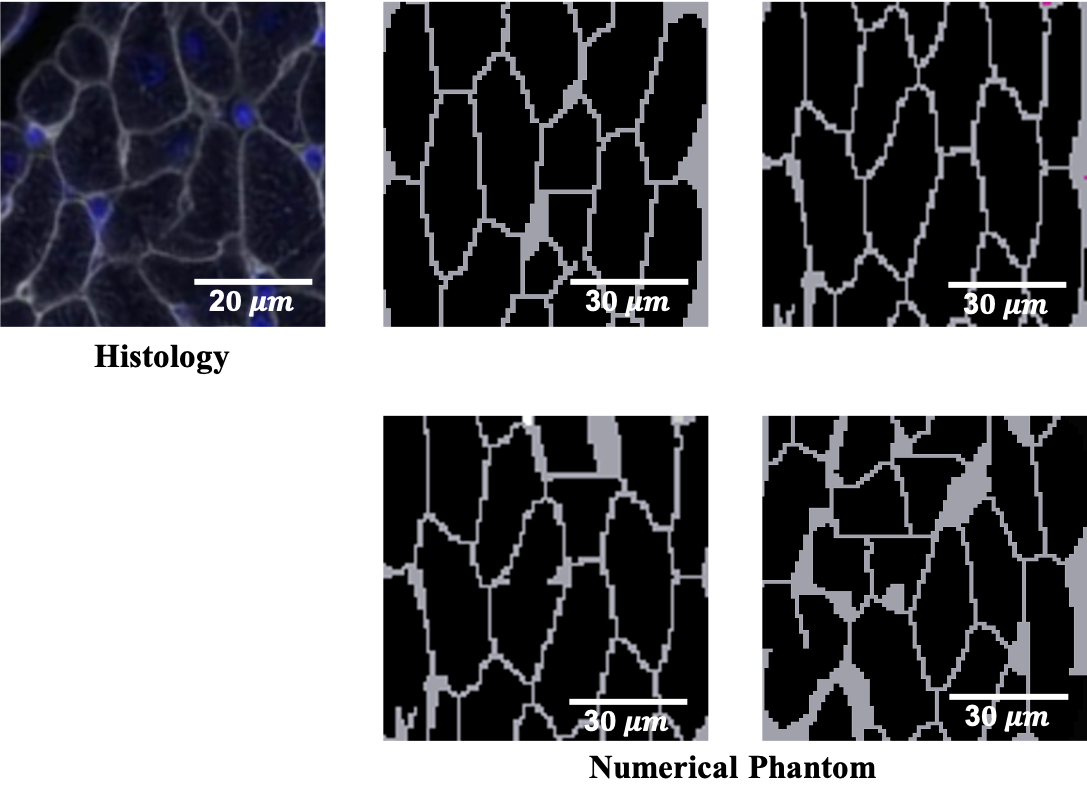}
         \caption{}
         \label{fig:CrossSection2}
     \end{subfigure}
     \hfill
     \begin{subfigure}[b]{0.4\textwidth}
         \centering
         \includegraphics[width=1\linewidth]{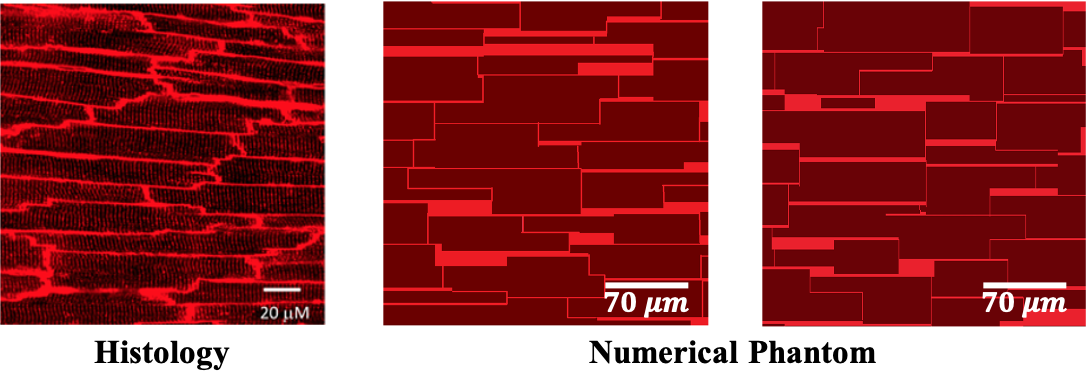}
         \caption{}
         \label{fig:LongCross}
     \end{subfigure} 
     \caption{Comparison between in-silico and real myocardial tissue. a) Comparison between EM image of single CMs \citep{bensley2016three} with in-silico CMs (the blues are microvasculature), b) Comparison between the transverse cross-section of myocardial tissue in a confocal microscopy image and an in-silico version of tissue, c) Comparison between the longitudinal cross-section \citep{chen2015situ} of myocardial tissue in a confocal microscopy image and an in-silico version of tissue.}
     \label{fig:VisualAssessment}
\end{figure}

This experiment aims to compare the in-silico version of: (i) single CMs, Figure \ref{fig:CMcomparison}; along with (ii) transverse cross-section, Figure \ref{fig:CrossSection2}; and (iii) longitudinal cross-section, Figure \ref{fig:LongCross}, of the myocardial tissue with their real counterparts using histological and confocal microscopy images.
Visual assessment of individual CMs and cross-sectional images from tissue histology and the in-silico phantom indicate good agreement between the former and latter.

\subsection{Morphometric Study of Virtual CMs} \label{sec:VB}
\begin{figure}[!tb]
\begin{center}
    \begin{subfigure}[normal]{.49\linewidth}
    \centering
    \includegraphics[width=1\linewidth]{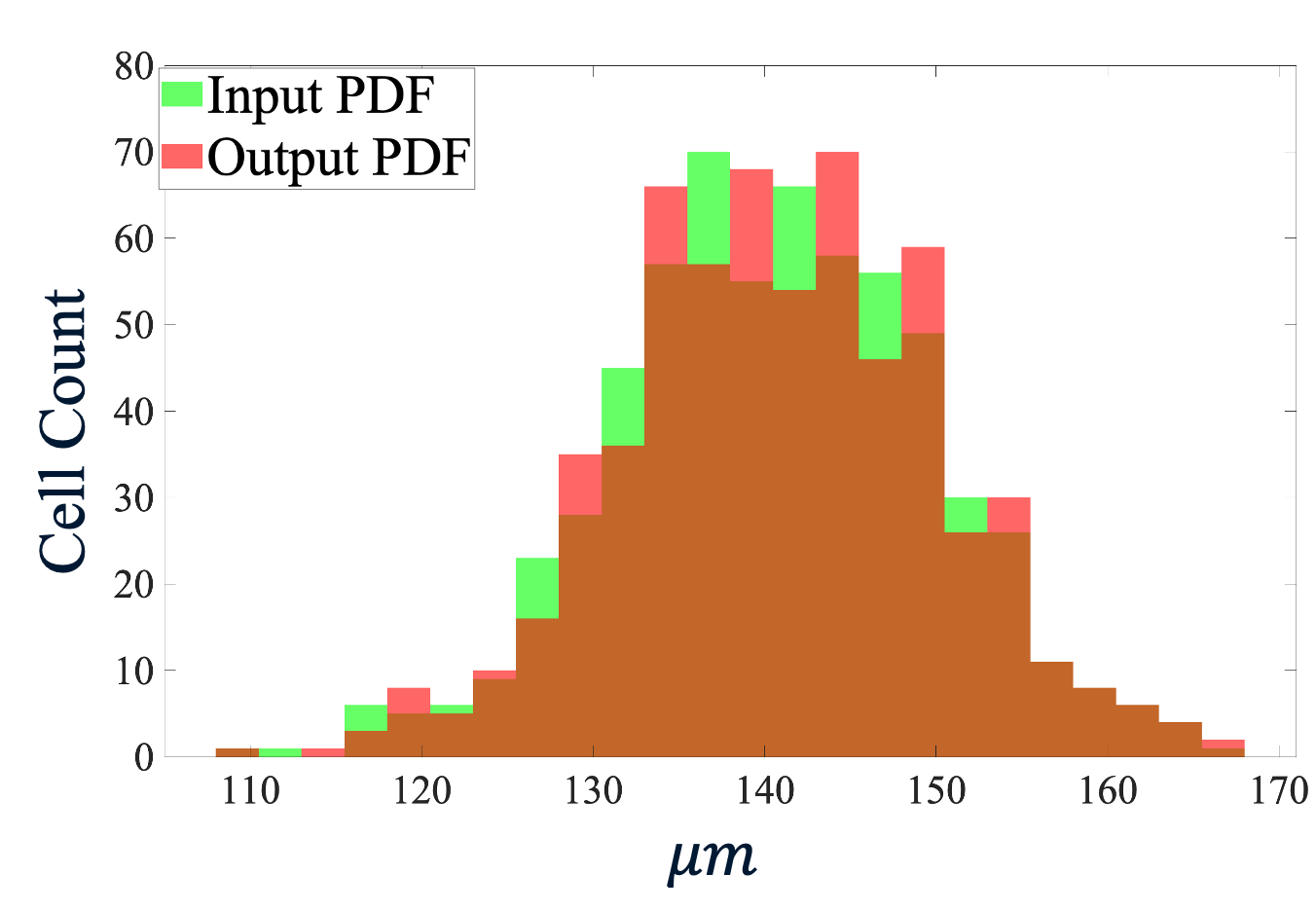}
    \caption{}
    \label{fig:LengthHist}
    \end{subfigure}
    \begin{subfigure}[normal]{.49\linewidth}
    \centering
    \includegraphics[width=1\linewidth]{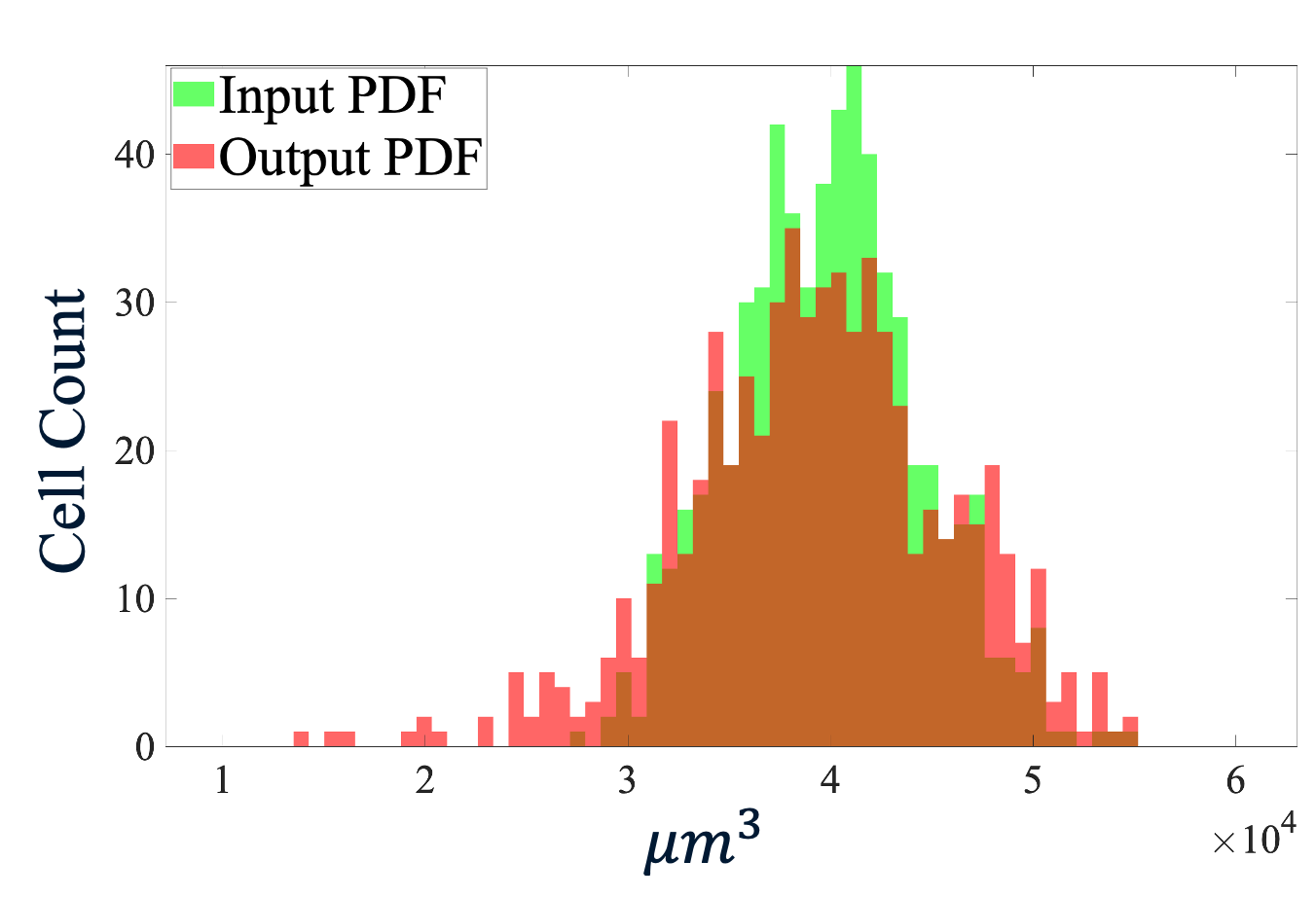}
    \caption{}
    \label{fig:VolumeHist}
    \end{subfigure}
    \begin{subfigure}[normal]{.49\linewidth}
    \centering
    \includegraphics[width=1\linewidth]{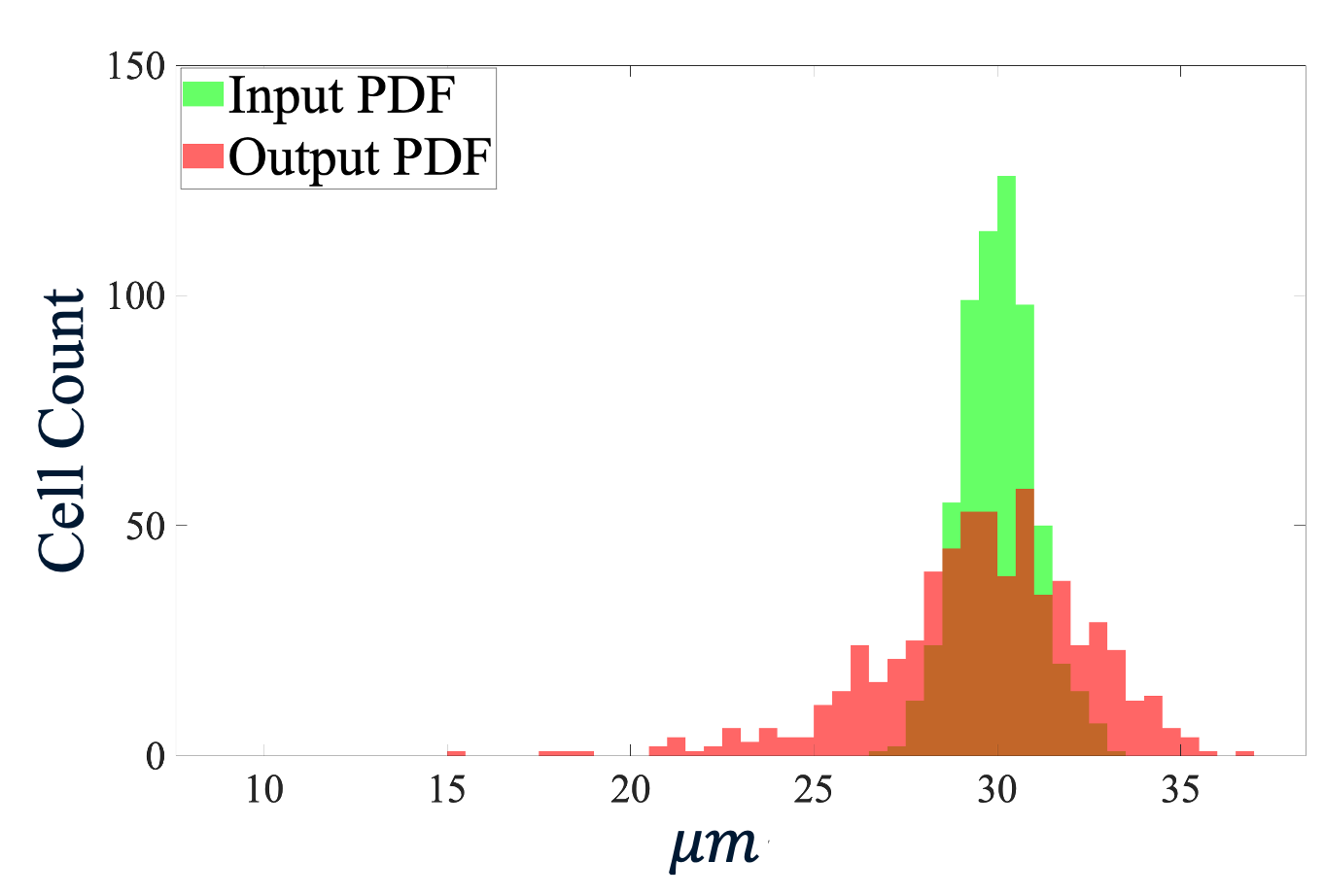}
    \caption{}
    \label{fig:MajorHist}
    \end{subfigure}
    \begin{subfigure}[normal]{.49\linewidth}
    \centering
    \includegraphics[width=1\linewidth]{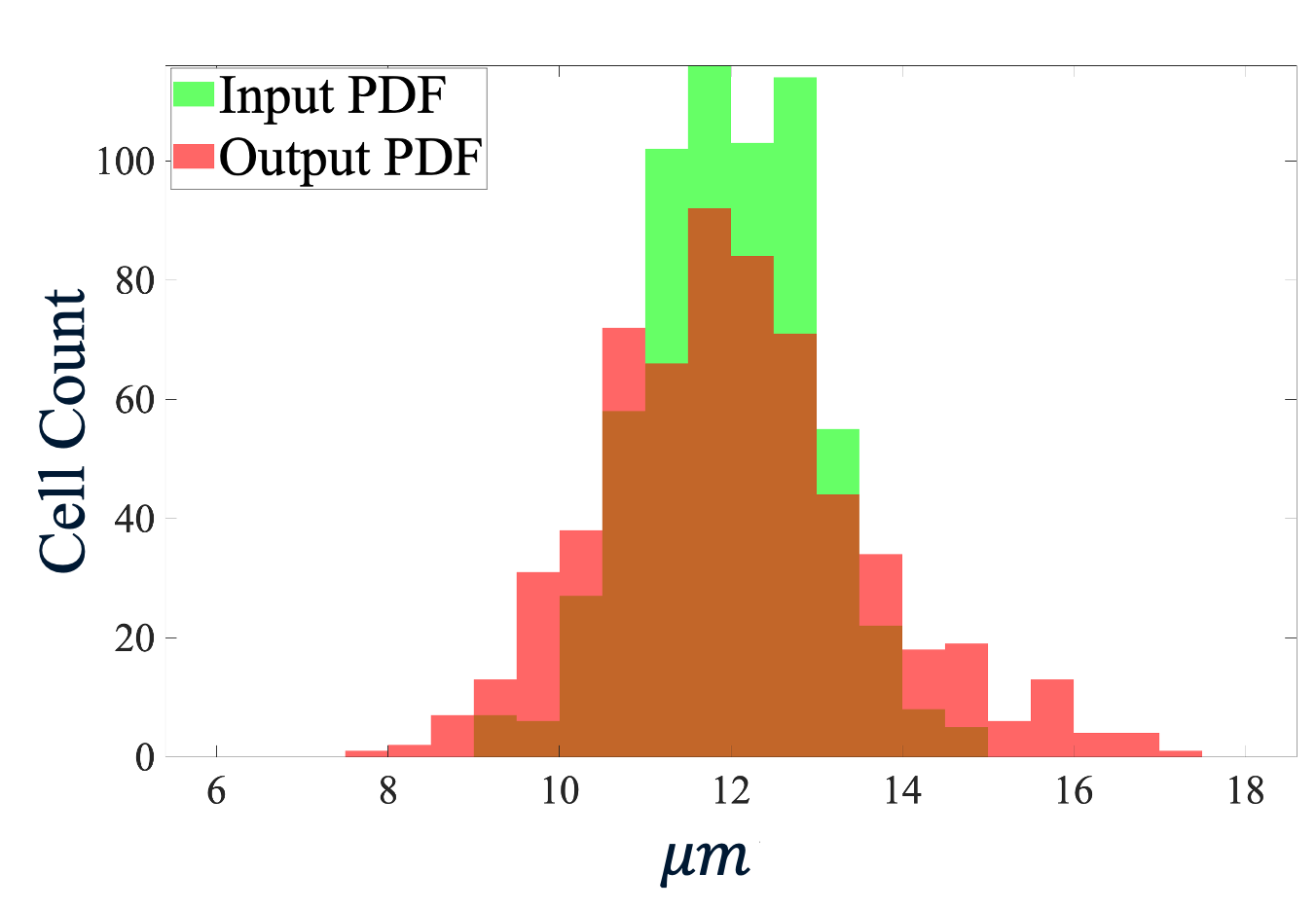}
    \caption{}
    \label{fig:MinorHist}
    \end{subfigure}
\end{center}
\caption{Comparison between the input and output PDFs of the CMs’ shape parameters. a) Length $L$, bin size of 2.5 $\mu m$ b) Volume $V$, bin size of 806 $\mu m^{3}$ c) Major-axis $A$, bin size of 0.5 $\mu m$ d) Minor-axis $B$, bin size of 0.5 $\mu m$.}
\label{fig:MyoHistMorphology}
\end{figure}

This experiment evaluates how the PDFs from literature, used as inputs, are preserved following phantom generation.
The virtual morphometric study is performed on 20 different numerical phantoms generated according to the PDFs shown in Table (\ref{t:Simpara}), each comprising 642 virtual CMs.
For each numerical phantom, all processes of phantom generation are repeated from the beginning.
The input PDFs are made on assumptions that the value reported (mean$\pm$standard deviation) in literature comes from a normal distribution.
For each numerical phantom, all processes of phantom generation are repeated from the beginning.
The volume $V$, length $L$, and profile $Area$ ($Area$ is defined as the maximum area of the 2D projection of virtual CMs on the plane parallel to the major-axis and perpendicular to the minor-axis of 2D polygons) are directly measured.
Then, CM's major-axis $A$ is calculated from $A=\frac{Area}{L}$, and consequently, CM's minor-axis $B$ is measured using microstructural restriction density, the volume formula for an elliptical cylinder, $V=\pi ABL$, as explained in \citep{chen2007method}.
Figure \ref{fig:MyoHistMorphology} illustrates an example of the input and output PDFs of the CMs' shape parameters.

\par In this experiment, two null hypotheses are tested: first, the output PDFs follow a normal distribution; second, there is no significant difference between the input and output PDFs.
The Shapiro-Wilk test tests the first null hypothesis, commonly used to test whether PDFs come from a normal distribution or not \citep{ahad2011sensitivity,nayak2011choose}.
The Mann—Whitney U-test tests the second null hypothesis if the first null hypothesis is rejected; otherwise, it is tested by the t-test.
Moreover, to compare the variance of the PDFs, the null hypothesis that the input and output PDFs have equal variances is tested using a two-sample F-test if both PDFs have a normal distribution, and a Brown-Forsythe test if one of the PDFs does not follow a normal distribution.
Table \ref{t:MyoMeanVarMorphology} presents these statistical tests and a comparison between the input and output data.
Data are expressed as mean $\pm$ SD and are compared using the Shapiro\textemdash Wilk test, Mann\textemdash Whitney U-test, t-test, two-sample F-test, and Brown\textemdash Forsythe test.
\par The resultant \textit{p} values for the Shapiro-Wilk test, \textit{p}$_{1}$-values in Table \ref{t:MyoMeanVarMorphology}, indicate that the proposed method preserves normality for the length and volume PDFs but does not for major- and minor-axis PDFs.
These results are coupled with instance histograms are shown in Figure \ref{fig:MyoHistMorphology}, where the length and volume of output PDFs (red bars in Figures \ref{fig:LengthHist} and \ref{fig:VolumeHist}) are symmetric around their means, whereas it does not hold for PDFs of the major- and minor-axis (red bars in Figures \ref{fig:MajorHist} and \ref{fig:MinorHist}).
Moreover, the null hypothesis of no statistically significant difference between input and output PDFs is confirmed based on the \textit{p}$_{2}$- and \textit{p}$_{3}$-values. 
Finally, according to \textit{p}$_{4}$- and \textit{p}$_{5}$-values, the null hypothesis of the equal variances for input and output PDFs only fails to reject for length PDF.

\begin{table*}[!tb]
\caption{Statistical tests and comparison between the input and output means} $\pm$ SD of the CMs’ shape parameters: Length $L$, Volume $V$, Major-axis $A$, and Minor-axis $B$.
    \centering
    \begin{tabular}{M{1cm}M{2cm}M{2.5cm}M{2cm}M{2cm}}
    \hline\hline
     & $L (\mu m)$ & $V (\mu m^3)$ & $A (\mu m)$ & $B (\mu m)$ \\ \hline
    Input & $141.1\pm9.3$ & $39,933\pm4,640$ & $30.1\pm1.1$ & $11.1\pm1$ \\
    Output & $141.1\pm9.4$ & $39,732\pm6,693$ & $30.2\pm2.9$ & $11.9\pm1.6$ \\
    \textit{p}$_{1}$ & $0.28>0.01$ & $0.17>0.01$ & $0<0.01$ & $0<0.01$ \\
    \textit{p}$_{2}$ & - & - & $0.09>0.01$ & $0.19>0.01$ \\
    \textit{p}$_{3}$ & $0.99>0.01$ & $0.76>0.01$ & - & - \\
    \textit{p}$_{4}$ & $0.98>0.01$ & $10^{-18}<0.01$ & - & - \\
    \textit{p}$_{5}$ & - & - & $10^{-53}<0.01$ & $10^{-18}<0.01$ \\
    \hline\hline
    \end{tabular}
    \begin{tablenotes}
        \item[1]Values are mean $\pm$ SD for 642 CMs from 20 phantoms.
        \item[2]\textit{p}$_{1}$, \textit{p}$_{2}$, \textit{p}$_{3}$, \textit{p}$_{4}$, and \textit{p}$_{5}$ are the results of the Shapiro\textemdash Wilk test, Mann\textemdash Whitney U-test, t-test, two-sample F-test, and Brown\textemdash Forsythe test, respectively.
        A value of $\textit{p} < 0.01$ is considered statistically significant.
    \end{tablenotes}
\label{t:MyoMeanVarMorphology}
\end{table*}

\subsection{Microstructural Complexity of the In-silico Tissue}
This experiment investigates the similarity between the structural universality classes of the in-silico myocardial tissue and the histological data from real tissue.
\cite{novikov2014revealing} classified the microstructure into different \textit{structural universality classes} based on the unique types of long-scale spatial correlations of a medium at the microscale. 
The criterion for this classification is the exponent $p$ in $\Gamma(k)|_{k\rightarrow0} \sim k^{p}$, which describes the long-range fluctuations of microstructural restriction density.
$\Gamma(k)$ is radially averaged of the power spectrum density ($PSD(u,v)$) of the tissue microstructure around $k=0$, where $k=\sqrt{u^2+v^2}$ \citep{lee2019revealing}.
According to Wiener–Khinchin theorem, $PSD(u,v)$ can be found using the Fourier transform of the autocorrelation function \citep{lathi1998modern}:
\begin{equation}\label{eq:StructuralCorrelation}
    PSD(u,v) = \sum_{x=0}^{M-1} \sum_{y=0}^{N-1} C(x,y) e^{-j2\pi(\frac{ux}{M} + \frac{vy}{N})}
\end{equation}
where $C(x,y)$ is the autocorrelation of a 2D image, $I(x,y)$, $0<x<M$, $0<y<N$.
\par
\begin{figure}[!tb]
\centering
\includegraphics[width=3.8in]{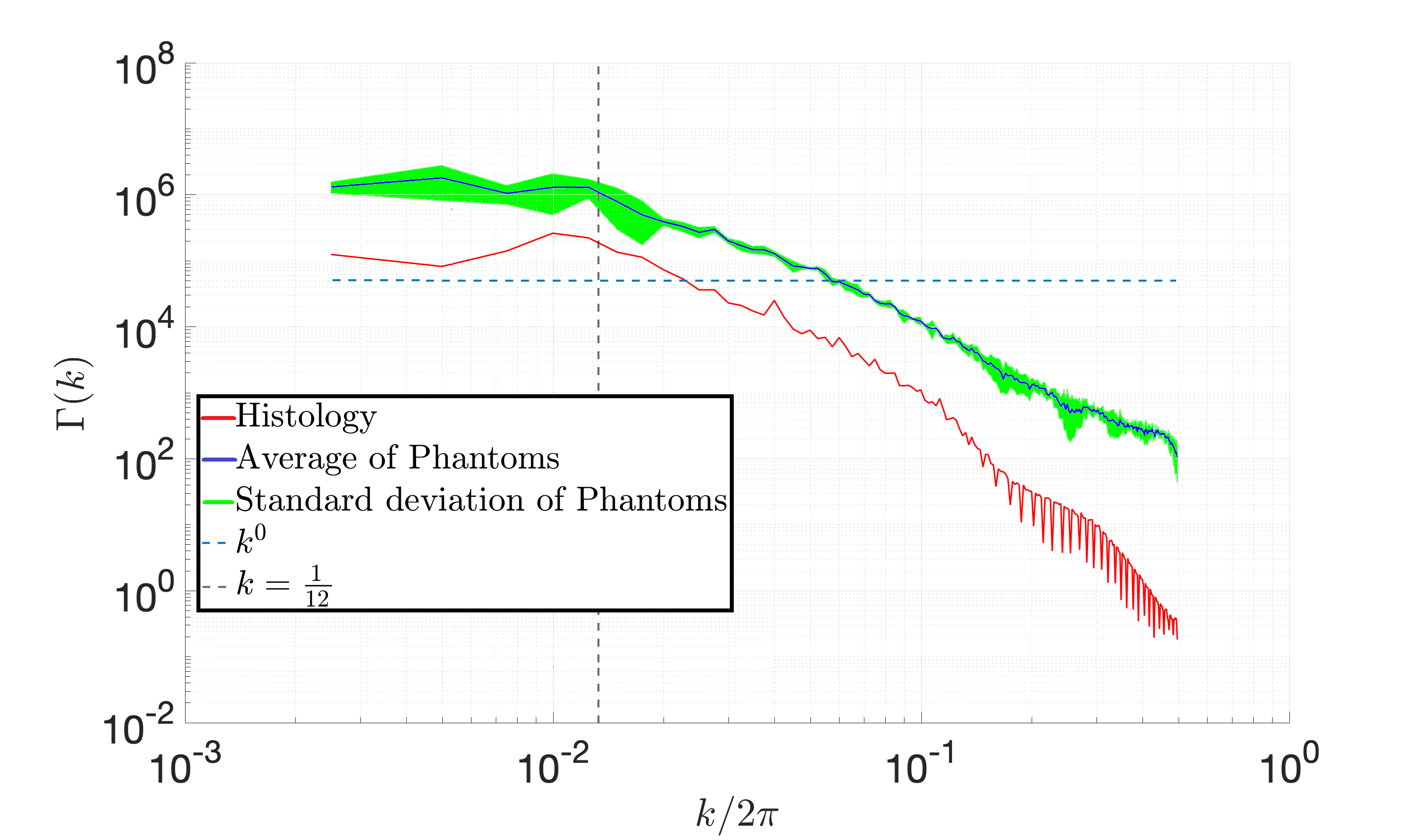}
\caption{$\Gamma(k)$ for 2D images of the cross-sections of the in-silico and real myocardial tissues shown in Figure \ref{fig:CrossSection2}.}
\label{fig:StructCooreCMs}
\end{figure}

 Here, we compare the value of $\Gamma(k)$ for the in-silico and histological images of the myocardial transverse cross-section (in the plane perpendicular to the long axis of CMs), shown in Figure \ref{fig:CrossSection2}. 
Figure \ref{fig:StructCooreCMs} depicts the diagrams of $\Gamma(k)$ for these images, at the different $k$.
As illustrated in Figure \ref{fig:StructCooreCMs}, at $k<\frac{1}{CM's\ minor-axis}$ i.e., $k<\frac{1}{12}$, the diagrams of the in-silico tissue follow the same power-law tail of the real tissue, i.e., $k^{0}$.

\subsection{cDTI MRI Measurements vs. Phantom-based Simulations} \label{sec:5.4}
\begin{figure}[!tb]
\centering
    \begin{subfigure}[normal]{.34\linewidth}
    \centering
    \includegraphics[width=1\linewidth]{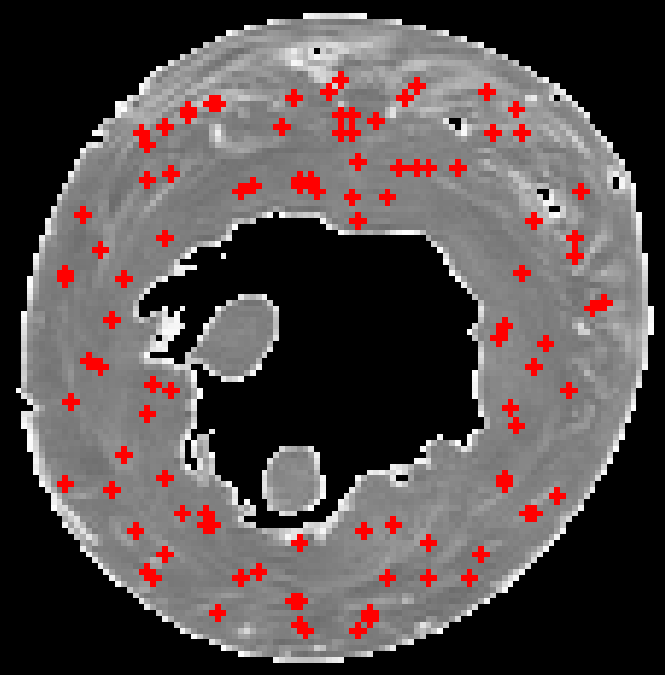}
    \caption{}
    \label{fig:VoxelLocation}
    \end{subfigure}
    \begin{subfigure}[normal]{.64\linewidth}
    \centering
    \includegraphics[width=1\linewidth]{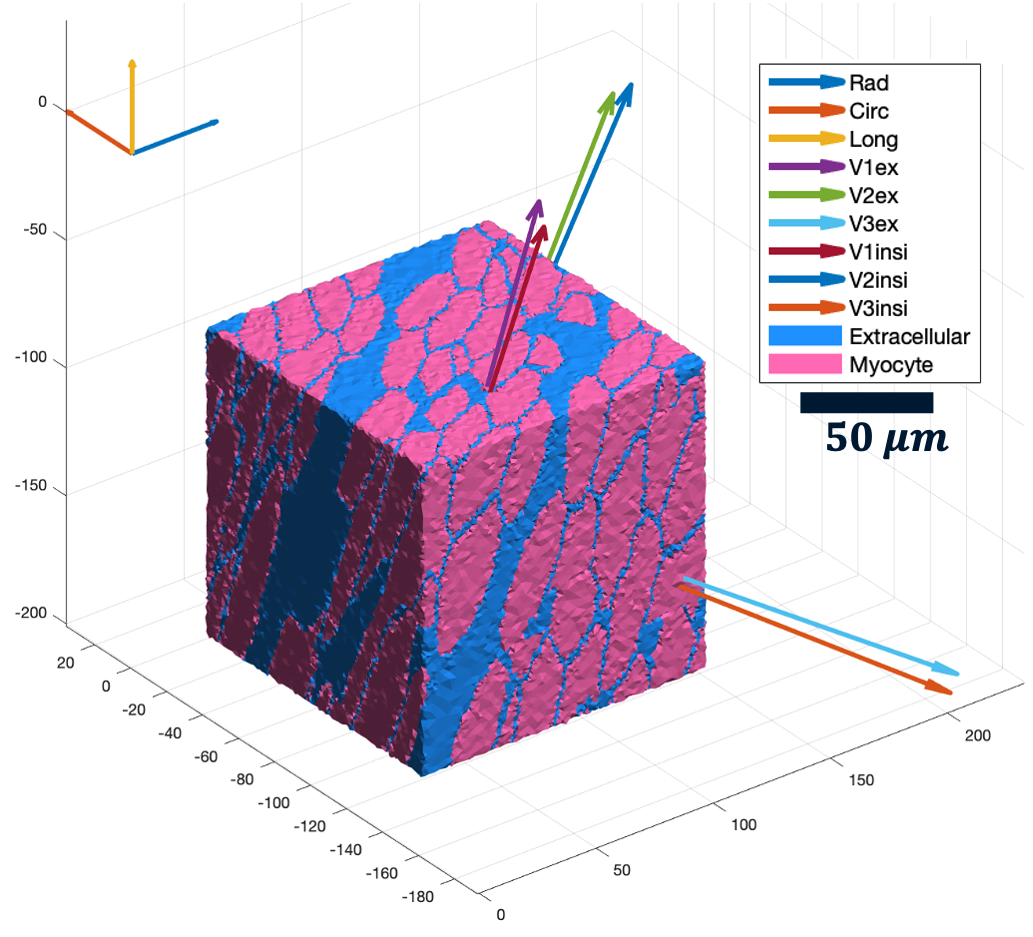}
    \caption{}
    \label{fig:VoxelSample}
    \end{subfigure}
    \caption{a) Location of mimicking voxels in a slice of experimental cDTI, b) An example of in-silico voxel oriented according to voxel 8 in (a).}
\end{figure}

This experiment investigates the performance of numerical phantom in replicating cDTI eigenvectors and eigenvalues using a simplified version of the numerical phantom.
The reason for this simplification is the lack of information about the chamber curvature and myocardial tissue twisting of the ex-vivo data.
Therefore, we discarded these features during the generation of the numerical phantom.
Moreover, collagen fibres are not included in the numerical phantom, since the ex-vivo cDTI is measured from a healthy heart, where the collagen VF is $\sim2\%$ \citep{haddad2017novel}.
Due to lack of information about inter-CMs space, this parameter was set $\SI{1}{\micro\meter}$.
\par cDTI eigenvectors correspond to CMs and sheetlets directions \citep{tseng2003diffusion,magat20213d}.
Therefore, we investigated how the simulated eigenvectors preserve the input directions.
First, we generated 100 in-silico voxels at the resolution of ex-vivo data, described in Section \ref{sec:data}, (the mesh resolution determined in Appendix \ref{appendix:MeshAnalysis}) and $VF_{ic}$ of the generated voxels fell into the range of 64\textemdash74\%, with 69\%$\pm$2\% (mean$\pm$SD).
Then, the directions of CMs and sheetlets of these voxels are oriented according to the eigenvectors of the voxels of an equivalent cDTI experiment, selected randomly from different myocardial regions (Figure \ref{fig:VoxelLocation}).
Finally, the biophysical parameters of the in-silico voxels are set up according to the information provided in Table \ref{t:Simpara}.

\begin{table}
\rotatebox{90}{\begin{minipage}{\textheight}
    \caption{Phantom parameters for healthy myocardium}
    \label{t:Simpara}
    \centering
    \begin{threeparttable}
    \begin{tabular}{M{1.2cm}M{2.4cm}M{2.4cm}M{2.4cm}M{3.3cm}M{2.4cm}M{2.4cm}M{2.2cm}M{2.2cm}}
    \hline\hline
    \rowcolor{lightgray}
     & $L(\mu m)$ & $V(\mu m^3)$ & $A(\mu m)$ & $B(\mu m)$ & $D_{ic}$($\SI{}{\micro\meter^2\per\milli\second}$) & $D_{ex}$($\SI{}{\micro\meter^2\per\milli\second}$) & $T_{2_{ic}}$($\SI{}{\milli\second}$) & $T_{2_{ex}}$($\SI{}{\milli\second}$) \\
    \hline
    Literature & $N(141.1\pm9.3)$ \citep{chen2007method} & $N(39933\pm4640)$ \citep{chen2007method} & $N(30.1\pm1.1)$ \citep{chen2007method} & $N(11.1\pm1)$ \citep{chen2007method} & 0.83 & 1.91 & 18.4 & 42.4 \\
    ex-vivo & NA & NA & NA & NA & NA & NA & NA & NA \\
     & & & & & & & & \\
    \hline
    \rowcolor{lightgray}
     & $\kappa_{Sarco.}$($\SI{}{\micro\meter\per\second}$) & $\kappa_{ICDs}$($\SI{}{\micro\meter\per\second}$) & $p_{r}$($\SI{}{\per\meter}$) & $p_{l}$($\SI{}{\per\meter}$) & $\alpha$(°) & \textbf{$V_1$}(°) & \textbf{$V_2$}(°) & \textbf{$V_3$}(°) \\
    \hline
    Literature & 15 \citep{bates2017monte} & 0.5 \citep{bastide1996effect} & NA & NA & NA & - & - & - \\
    ex-vivo & NA & NA & NA & NA & NA & AVL & AVL & AVL \\
    \hline
    \rowcolor{lightgray}
    & Sheetlet thickness ($\SI{}{\micro\meter}$) & inter-CMs space ($\SI{}{\micro\meter}$) & inter-sheetlet space ($\SI{}{\micro\meter}$) & $VF_{ic}$ (\%) & $VF_{collagen}$ (\%) & $D_{collagen}$ ($\SI{}{\micro\meter^2\per\milli\second}$) & $T_{2_{collagen}}$ ($\SI{}{\micro\meter\per\second}$) & \\
    \hline
    Literature & $\sim$2\textemdash4 CMs thick \citep{hales2012histo}. & NA & $\sim$1\textemdash2 CMs thick \citep{legrice2005architecture} & 65\textemdash75\% \citep{greiner2018confocal, skepper1995ultrastructural} & $\sim2\%$ \citep{haddad2017novel} & NA & NA & \\
    ex-vivo & NA & NA & NA & NA & NA & NA & NA & \\
    \hline\hline
    \end{tabular}
    \begin{tablenotes}
        \item[] AVL: available, NA: not available.
    \end{tablenotes}
\end{threeparttable}
\end{minipage}}
\end{table}

As shown in Table \ref{t:Simpara}, the eigenvectors of in-silico voxels are exactly matched to the eigenvectors of the ex-vivo data.
$T_{2_{ex}}$ is set $\SI{42.4}{\milli\second}$, water $T_{2}$ measured in the field strength of 9.4 T \citep{lei2003changes}.
Then, $T_{2_{ic}}$ is computed by assuming $T_{2_{ic}}\times VF_{ic} + T_{2_{ex}}\times VF_{ex} = T_{2_{ex-vivo}}$ where $T_{2_{ex-vivo}} = \SI{25.72}{\milli\second}$ is the average value of $T_{2}$ for 100 selected ex-vivo voxels.
Similarly, $D_{ex}$ is set $\SI{1.91}{\micro\meter^2\per\milli\second}$, free water diffusivity reported in \citep{periquito2019diffusion}, due to similar temperature during the acquisition in \citep{periquito2019diffusion} and the ex-vivo data under comparison \citep{teh2016resolving}.
Then, $D_{ic}$ is computed by assuming $D_{ic}\times VF_{ic} + D_{ex}\times VF_{ex} = D_{ex-vivo}$ where $D_{ex-vivo} = \SI{1.06}{\micro\meter^2\per\milli\second}$ is the average diffusivity for 100 selected ex-vivo voxels.
Since $D_{ex-vivo}$ is reduced by CMs’ geometry, the computed $D_{ic}$ needs to increase to offset the effect of CMs’ geometry. 
Due to a lack of quantitative information about diffusivity reduction caused by CMs’ geometry, different values of $D_{ic}$ are assessed to achieve the best possible agreement between eigenvalues of the in-silico and ex-vivo measurements for 16 voxels, chosen randomly. 
According to this assessment, the best possible $D_{ic}$ is achieved by 20\% increase i.e., $\!\SI{0.83}{\micro\meter^2\per\milli\second}$, and set in the simulation for the remaining voxels.

Other input parameters used for cDTI simulations were taken from published ex-vivo data in the literature.
Afterward, dMRI signal is simulated for 100 in-silico voxels by setting the imaging parameters of simulation such as b-values, diffusion encoding directions, diffusion time, diffusion encoding gradient type and etc. as same as ex-vivo data, described in Section \ref{sec:data}.
Finally, for each in-silico voxel, the eigenvectors, eigenvalues, and FA of cDTI are computed and compared with their counterpart from ex-vivo data.

\begin{figure}[!b]
\centering
    \begin{subfigure}[normal]{.45\linewidth}
        \centering
        \includegraphics[width=1\textwidth]{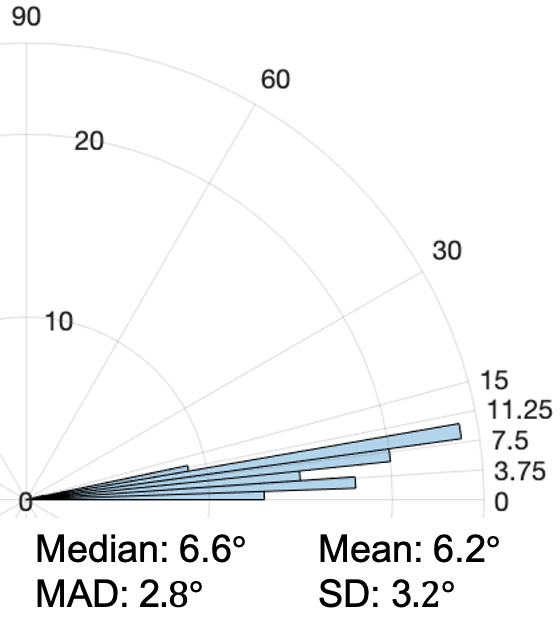}
        \caption{}
        \label{fig:V1}
    \end{subfigure}
    \hfill
    \begin{subfigure}[normal]{.45\linewidth}
        \centering
        \includegraphics[width=1\textwidth]{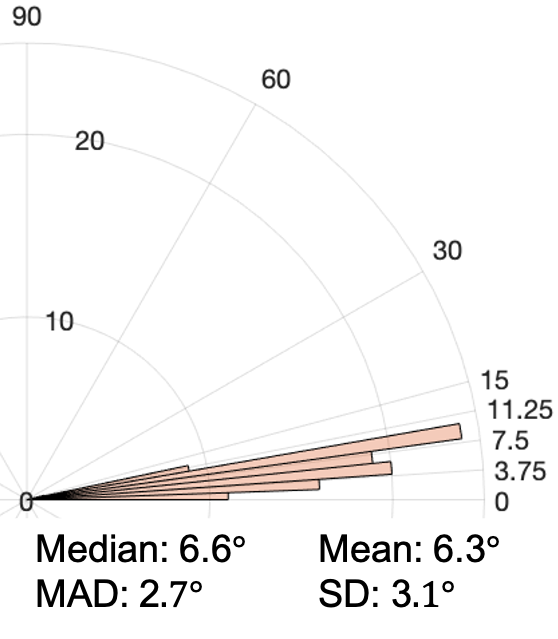}
        \caption{}
        \label{fig:V2}
    \end{subfigure}  
    \hfill
    \begin{subfigure}[normal]{.45\linewidth}
        \centering
        \includegraphics[width=1\textwidth]{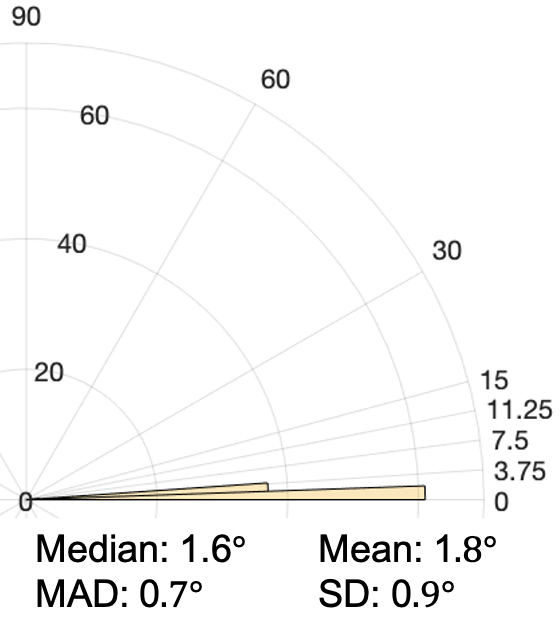}
        \caption{}
        \label{fig:V3}
    \end{subfigure}    
\caption{Angular distance between a) $V_1$, b) $V_2$, and c) $V_3$ of ex-vivo and in-silico voxels.}
\label{fig:AngleDiffV}
\end{figure}

\begin{figure}[!tb]
\centering
    \begin{subfigure}[normal]{.45\linewidth}
        \centering
        \includegraphics[width=1\textwidth]{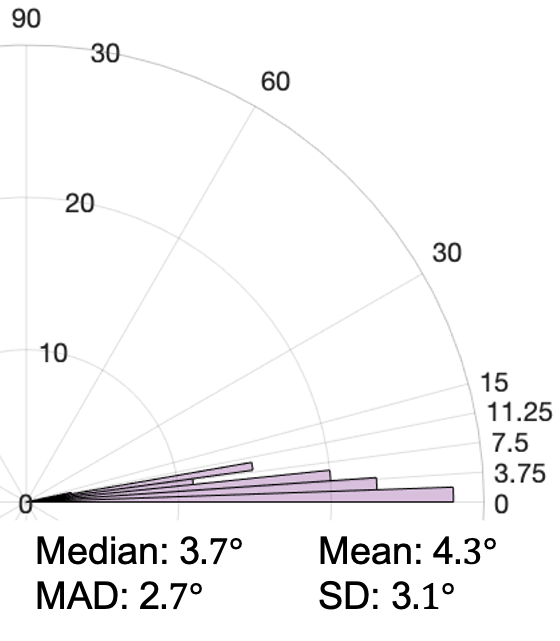}
        \caption{}
        \label{fig:HA}
    \end{subfigure}
    \hfill
    \begin{subfigure}[normal]{.45\linewidth}
        \centering
        \includegraphics[width=1\textwidth]{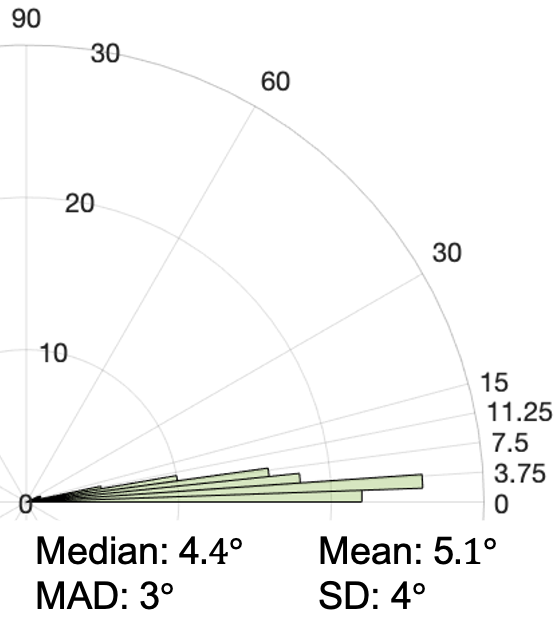}
        \caption{}
        \label{fig:TA}
    \end{subfigure}   
    \hfill
    \begin{subfigure}[normal]{.45\linewidth}
        \centering
        \includegraphics[width=1\textwidth]{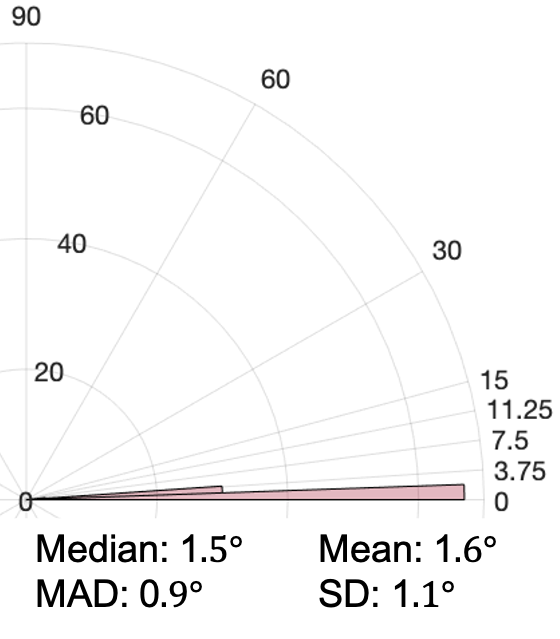}
        \caption{}
        \label{fig:SA}
    \end{subfigure}
    \hfill
    \begin{subfigure}[normal]{.45\linewidth}
        \centering
        \includegraphics[width=1\textwidth]{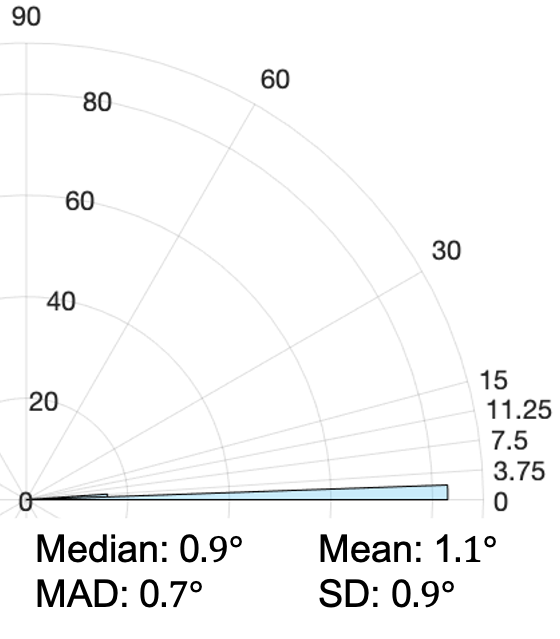}
        \caption{}
        \label{fig:SE}
    \end{subfigure}  
\caption{Absolute angle difference between a) HA, b) TA, c) SA, and d) SE of ex-vivo and in-silico voxels.}
\label{fig:AngleDiffS}
\end{figure}

\begin{figure*}[t]
\begin{center}
    \begin{subfigure}[normal]{.34\linewidth}
    \centering
    \includegraphics[width=1\linewidth]{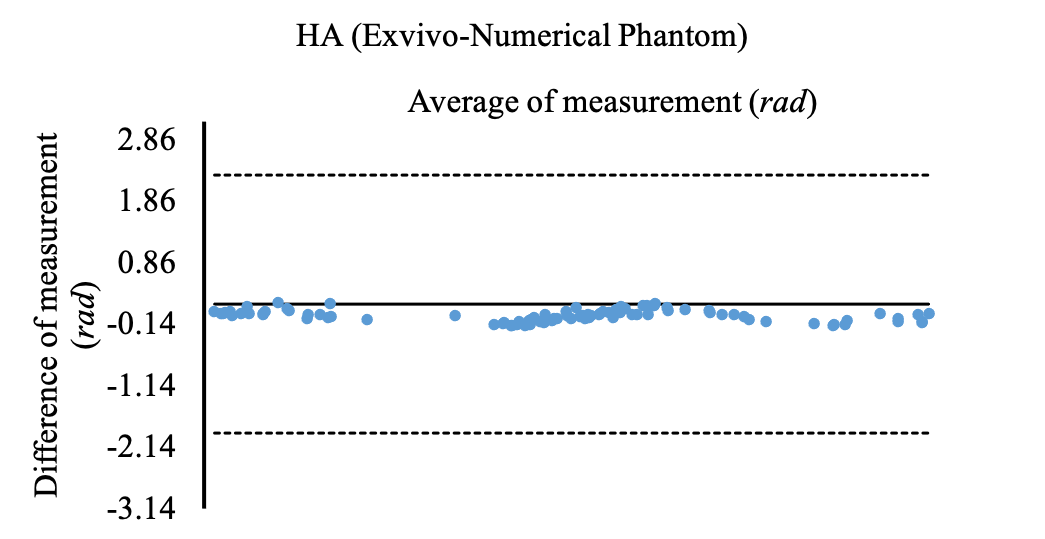}
    \caption{}
    \label{fig:BAE1}
    \end{subfigure}
    \hspace{1cm}
    \vspace{0.5cm}
    \begin{subfigure}[normal]{.34\linewidth}
    \centering
    \includegraphics[width=1\linewidth]{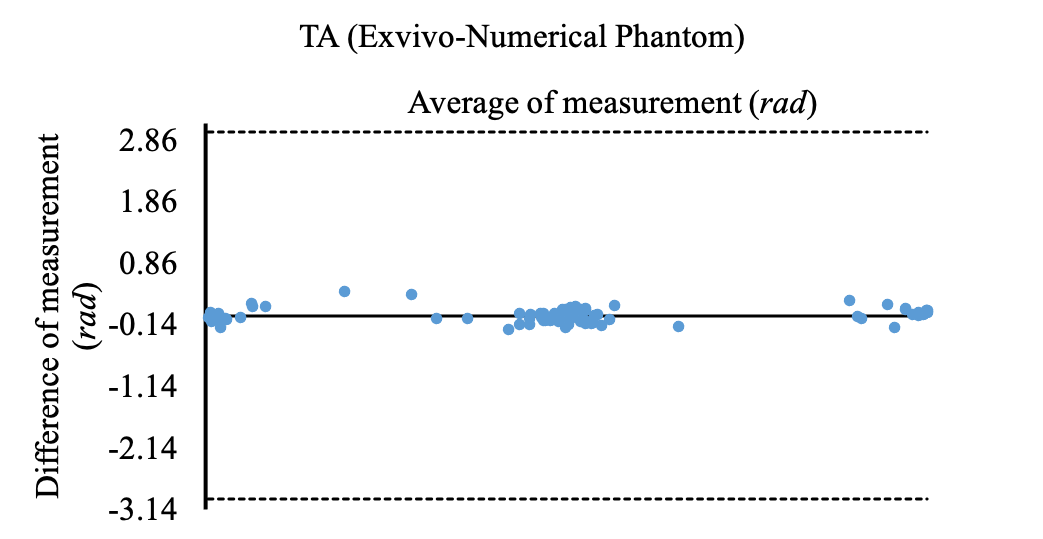}
    \caption{}
    \label{fig:BAE2}
    \end{subfigure}
    \vspace{0.5cm}
    \begin{subfigure}[normal]{.34\linewidth}
    \centering
    \includegraphics[width=1\linewidth]{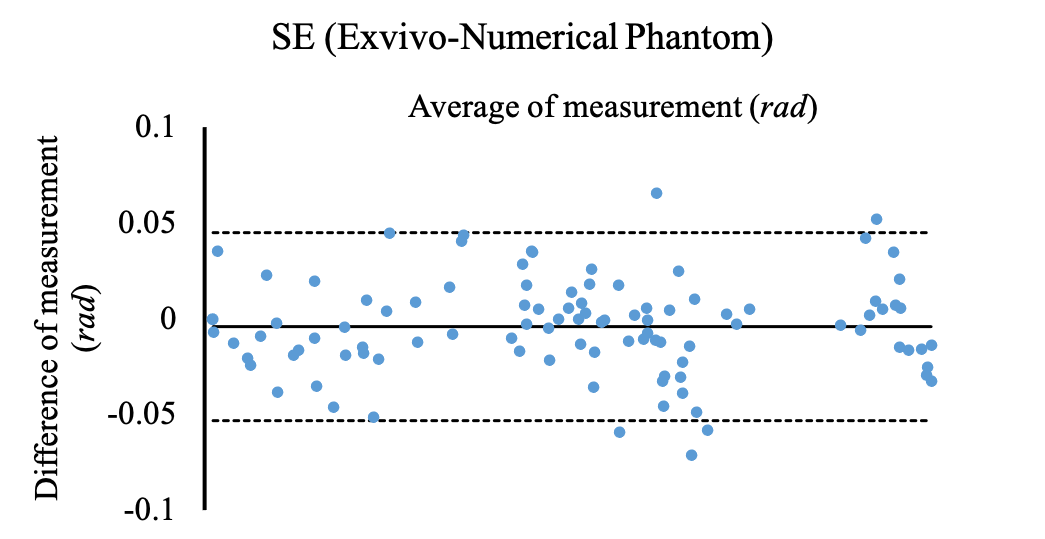}
    \caption{}
    \label{fig:BAE3}
    \end{subfigure}
    \hspace{1cm}
    \begin{subfigure}[normal]{.34\linewidth}
    \centering
    \includegraphics[width=1\linewidth]{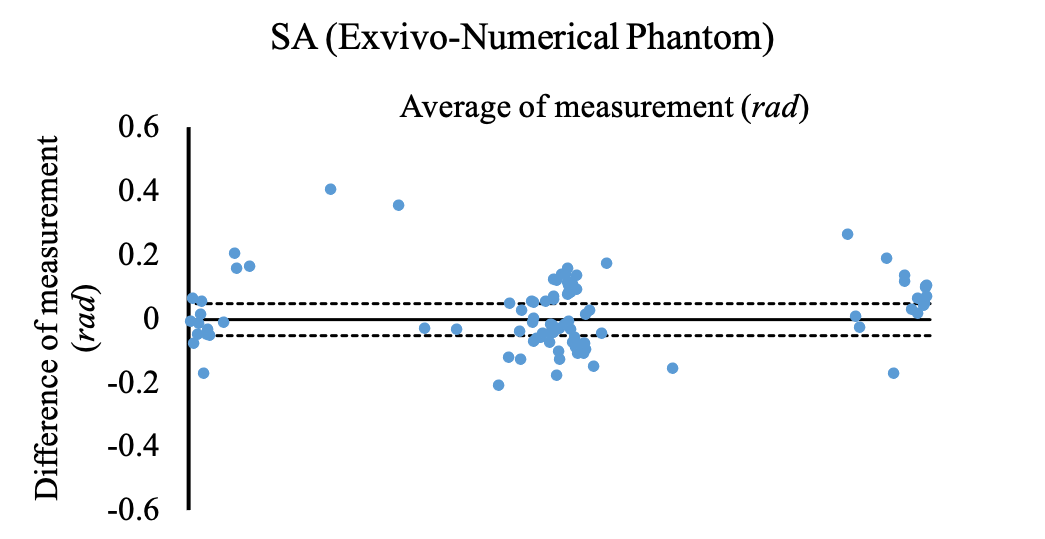}
    \caption{}
    \label{fig:BAE4}
    \end{subfigure}
\end{center}
\caption{a) Comparison between sheetlet angles of the experimental data and numerical phantom. Agreement between a) HA; b) TA; c) SE and d) SA of the experimental data and numerical phantom.}
\label{fig:BA}
\end{figure*}

Figures \ref{fig:AngleDiffV} and \ref{fig:AngleDiffS} show the distribution of the angular distance between eigenvectors, along with the absolute angle difference between HA, TA, SA, and SE of the input and simulated ones in rose diagrams.
The mean and standard deviation (SD), along with median and median absolute deviation (MAD) for these distributions are reported under each diagram.
These angular distance and absolute angle differences are much smaller than those reported between cDTI and structural tensor imaging (STI)  \citep{bernus2015comparison} shown in Table \ref{t:EigenvecDiff}, or reported in \citep{haliot20193d}.
In addition, as illustrated in the Bland-Altman plots in Figures \ref{fig:BAE1}, \ref{fig:BAE2}, \ref{fig:BAE3}, and \ref{fig:BAE4}, there is an adequate agreement between HA, TA, SE and SA of the ex-vivo data and the numerical phantoms.

 \begin{table*}
 \caption{Angular distance and absolute angle difference (median$\pm$MAD) between eigenvectors and sheetlet angles of ex-vivo and in-silico voxels.}
     \begin{center}
     \begin{tabular}{M{0.7cm}M{3.9cm}M{1.5cm}M{1.5cm}M{1.5cm}M{1.5cm}M{1.5cm}M{1.5cm}M{1.5cm}}
     \hline\hline
     & & \textbf{$V_1^\circ$}  & \textbf{$V_2^\circ$} & \textbf{$V_3^\circ$} & \textbf{$HA^\circ$}  & \textbf{$TA^\circ$} & \textbf{$SE^\circ$} & \textbf{$SA^\circ$} \\ \hline
     \multirow{2}{*}{Septal} & $\angle$input vs. $\angle$in-silico DTI & $7.1\pm3.1$ & $7.3\pm2.9$ & $1.3\pm0.7$ & $5.8\pm3.1$ & $3\pm3.9$ & $0.6\pm0.6$ & $1.1\pm0.9$ \\
     & $\angle$STI vs. $\angle$DTI & $13.3\pm6.7$ & $32.9\pm19.6$ & $27.9\pm17.4$ & $8.5\pm5.6$ & $11.5\pm7.8$ & $22\pm16.2$ & $22.9\pm16.5$ \\
     \hline
     \multirow{2}{*}{Lateral} & $\angle$input vs. $\angle$in-silico DTI & $4.7\pm3.1$ & $4.9\pm3$ & $2.2\pm0.8$ & $2.4\pm3.1$ & $4.4\pm1.7$ & $1.2\pm0.9$ & $1.7\pm0.9$ \\
     & $\angle$STI vs. $\angle$DTI & $12.6\pm5.9$ & $15\pm8.5$ & $23.8\pm11.9$ & $9.1\pm5.8$ & $15.6\pm11.1$ & $16.1\pm10.5$ & $14.6\pm10.3$ \\
     \hline\hline
     \end{tabular}
     \end{center}
 \label{t:EigenvecDiff} 
 \end{table*}

Additionally, the simulated eigenvalues, mean diffusivity (MD), fractional anisotropy (FA), and radial diffusivity (RD) are shown along with the eigenvalues of the ex-vivo data in Figure \ref{fig:Box}.
For these parameters two null hypotheses are tested: first, the distribution of the cDTI parameters for the experimental and simulation follows a normal distribution; second, there is no significant difference between the distribution of cDTI parameters of the experimental and simulated data.
These hypotheses are tested as in Section \ref{sec:VB} and their results are presented in Figure \ref{fig:Box} above each box and whisker plot.

\subsubsection{Computational cost}
The computations for Section \ref{sec:5.4} simulations were performed on ARC3, the High-Performance Computing facilities at the University of Leeds.
ARC3 consists of 252 nodes with 24 cores (Broadwell E5-2650v4 CPUs, 2.2 GHz) and 128GB of memory each and an SSD within the node with 100GB of storage. 
The details of computational cost for undertaken simulations in Section \ref{sec:5.4} is reported in Table \ref{t:CompCost}.

\begin{table*}[!tb]
    \caption{Computational cost of the simulation}
    \centering
    \begin{tabular}{M{3.1cm}M{2.3cm}M{1.8cm}M{3cm}M{2.1cm}M{3.7cm}} 
    \hline\hline
    Average number of tetrahedrons per voxel & Number of cores per voxel & Memory per core (GB) & Computational time per core (hours) & Entire Number of cores & Computational time for entire voxels (core hours) \\ 
    \hline
    856570$\pm$48666 & 7 & 62 & 6$\pm$2.5 & 700 & 4152 \\ 
    \hline\hline
    \end{tabular}
\label{t:CompCost} 
\end{table*}

\subsection{Effect of collagen density, twisting, and bending on MD and FA}
As mentioned earlier, due to insufficient information about the twisting and curvature of the myocardial wall along with the lack of information about the diffusivity and relaxation of collagenous ECS, these structures are excluded in the simulation presented in Section \ref{sec:5.4}.
The goal of the following experiment is to identify the effect of these structures on cDTI derivatives individually.
To this effect, we simulate cDTI for each structure's different range of values.
Increasing collagen density decreases diffusivity and relaxation \citep{mewton2011assessment,bun2012value,loganathan2006characterization}.
Figures \ref{fig:MDcolla} and \ref{fig:FAcolla} show that a decrease in diffusivity and relaxation values leads to a reduction in MD and FA. 
Here, we illustrate the effect of these changes on simulated signals for four pairs of values of diffusivities, $D_{collagen}\in [D_{ex}, D_{in}]$, and relaxations, $T_{2_{collagen}}\in [T_{2_{ex}}, T_{2_{in}}]$, respectively.
\par The degree of wall twisting and curvature depends on the studied cardiac phase. 
In the transition from systole to diastole, the degree of the twisting increases \citep{streeter1969fibre}, whereas the curvature decreases. 
To show the effect of these geometrical changes on MD and FA, cDTI is simulated using an in-silico voxel, mimicking transiting from systole to diastole at four points, where $\alpha^{\circ}\in [0, 60]$ and $p_l,p_r\in [\SI{0.31}{\per\milli\meter}, \SI{0.01}{\per\milli\meter}]$ \citep{ferferieva2018serial}, respectively.
Figures \ref{fig:MDtw} and \ref{fig:FAtw}; and Figures \ref{fig:MDcur} and \ref{fig:FAcur} show the simulated values of MD and FA, respectively, increase by decreasing the curvature and increasing the twisting, which agrees with changes in the in-vivo measurement of MD and FA \citep{khalique2020diffusion,mcgill2014comparison}.
The observation in Figure \ref{fig:FAtw} is somewhat surprising as FA values increase by increase of twisting, as it is expected that FA values decrease with increase in the range of CMs orientation resulting from increased twisting.
One possible explanation for this observation is that the increase in the tissue twisting leads to a decrease in CMs’ diameters \citep{axel2014probing, nielles2017assessment} which increases FA values.
The same reason may explain why increase in tissue bending results in increasing FA values (Figure \ref{fig:FAcur}).

\begin{figure*}[ht]
\begin{center}
    \begin{subfigure}[normal]{.25\linewidth}
    \centering
    \includegraphics[width=1\linewidth]{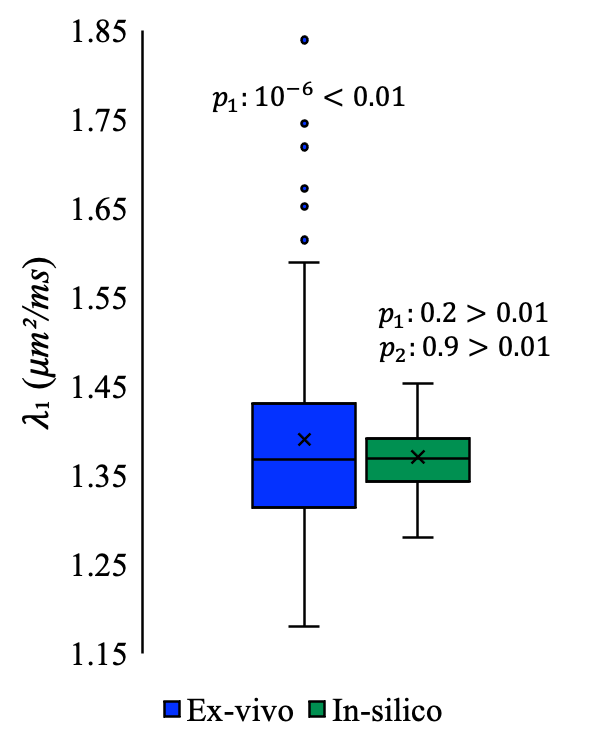}
    \caption{}
    \label{fig:eigenvalue1}
    \end{subfigure}
    \begin{subfigure}[normal]{.25\linewidth}
    \centering
    \includegraphics[width=1\linewidth]{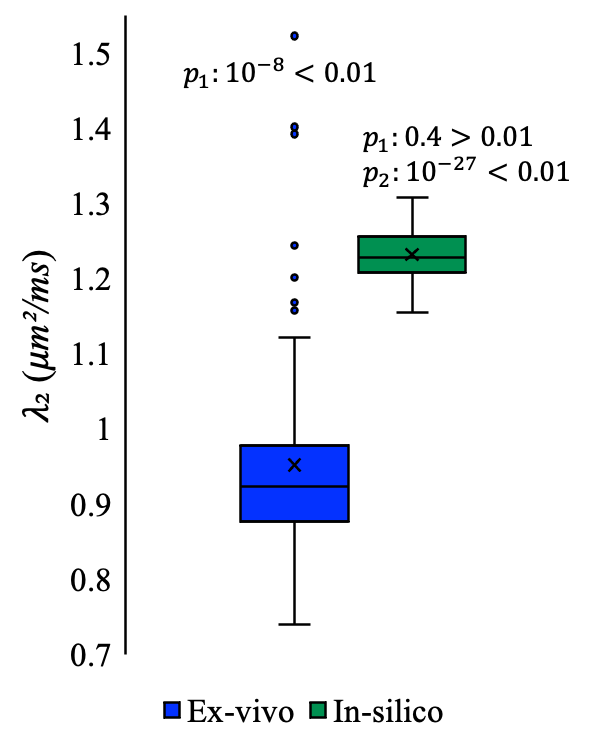}
    \caption{}
    \label{fig:eigenvalue2}
    \end{subfigure}
    \begin{subfigure}[normal]{.25\linewidth}
    \centering
    \includegraphics[width=1\linewidth]{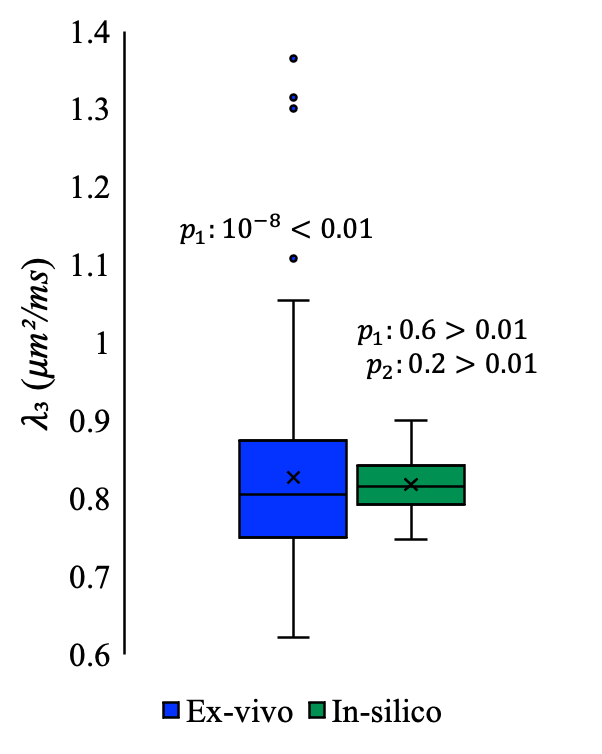}
    \caption{}
    \label{fig:eigenvalue3}
    \end{subfigure}    
    \begin{subfigure}[normal]{.25\linewidth}
    \centering
    \includegraphics[width=1\linewidth]{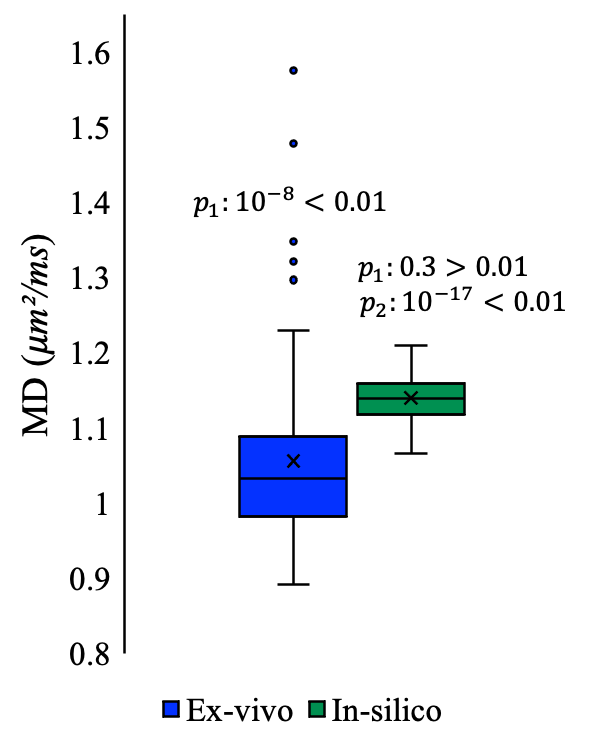}
    \caption{}
    \label{fig:BoxMD}
    \end{subfigure}
    \begin{subfigure}[normal]{.25\linewidth}
    \centering
    \includegraphics[width=1\linewidth]{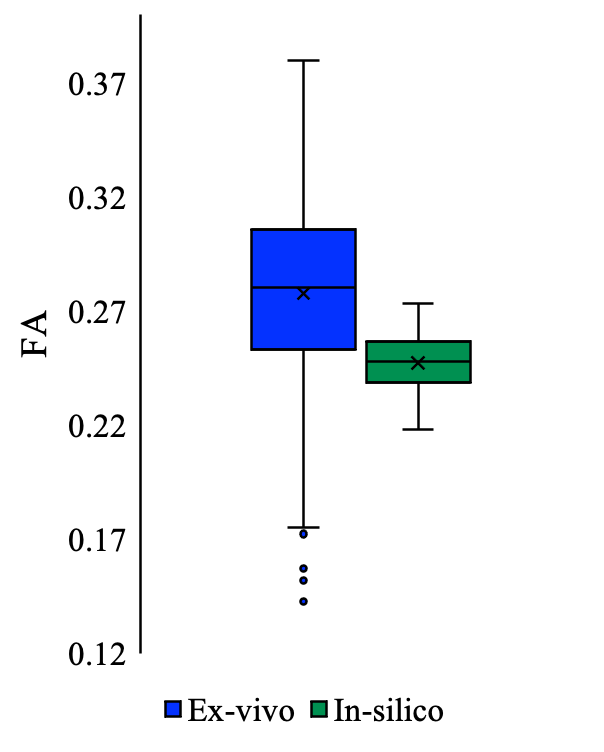}
    \caption{}
    \label{fig:BoxFA}
    \end{subfigure}
    \begin{subfigure}[normal]{.25\linewidth}
    \centering
    \includegraphics[width=1\linewidth]{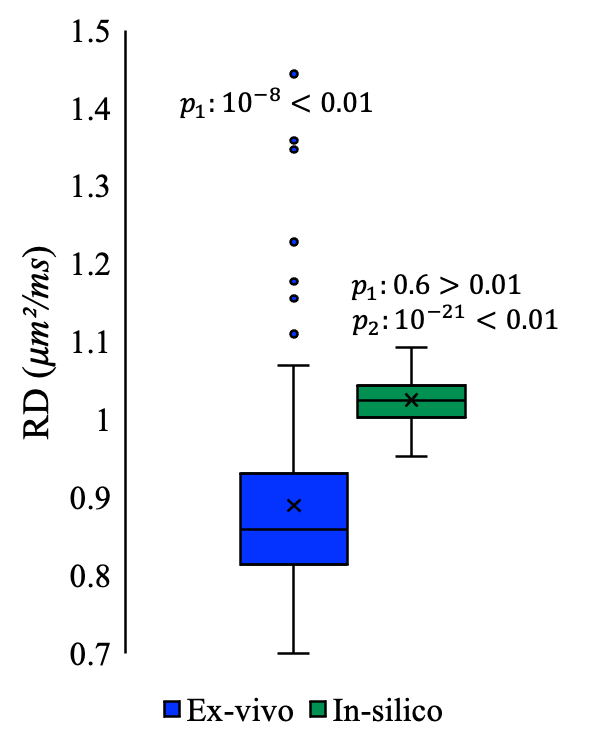}
    \caption{}
    \label{fig:BoxRD}
    \end{subfigure}
\end{center}
\caption{Comparison between cDTI parameters of the experimental (blue whisker plots) and in-silico data (green whisker plots). Agreement between a) $\lambda_1$; b) $\lambda_2$; c) $\lambda_3$; d) MD; e) FA; f) RD of the experimental data and numerical phantom.}
\label{fig:Box}
\end{figure*}

\begin{figure*}[ht]
\begin{center}
    \begin{subfigure}[normal]{.32\linewidth}
    \centering
    \includegraphics[width=1\linewidth]{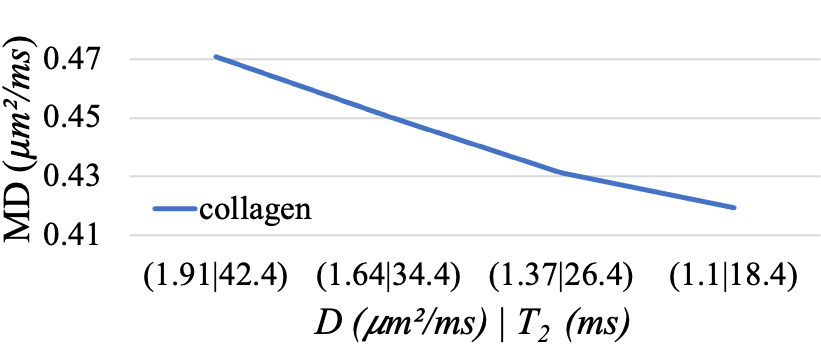}
    \caption{}
    \label{fig:MDcolla}
    \end{subfigure}
    \hspace{1cm}
    \begin{subfigure}[normal]{.25\linewidth}
    \centering
    \includegraphics[width=1\linewidth]{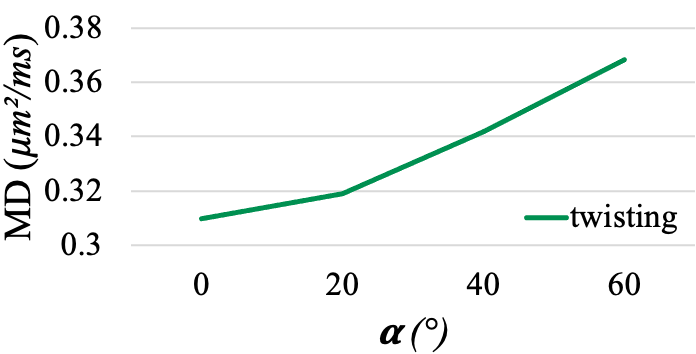}
    \caption{}
    \label{fig:MDtw}
    \end{subfigure}
    \hspace{1cm}
    \begin{subfigure}[normal]{.25\linewidth}
    \centering
    \includegraphics[width=1\linewidth]{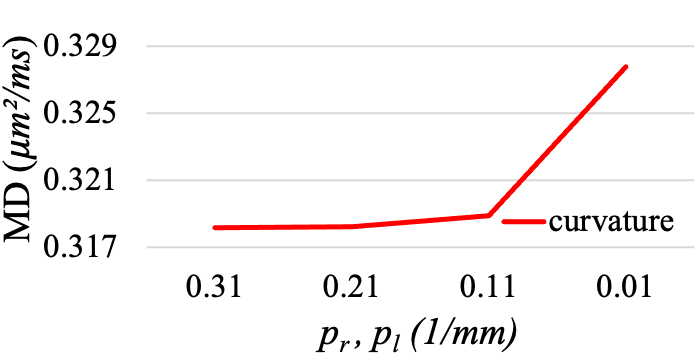}
    \caption{}
    \label{fig:MDcur}
    \end{subfigure}    
\end{center}
\caption{Effect of a) increase in collagen density, b) increase in twisting, and c) decrease in curvature on MD.}
\label{fig:MDCOllaTwCur}
\end{figure*}

\begin{figure*}[ht]
\begin{center}
    \begin{subfigure}[normal]{.32\linewidth}
    \centering
    \includegraphics[width=1\linewidth]{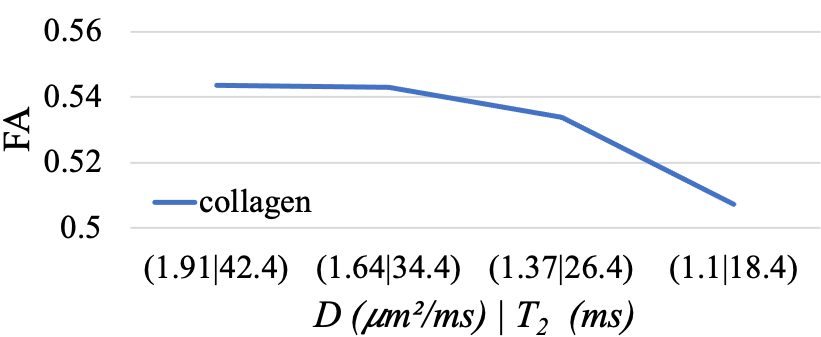}
    \caption{}
    \label{fig:FAcolla}
    \end{subfigure}
    \hspace{1cm}
    \begin{subfigure}[normal]{.25\linewidth}
    \centering
    \includegraphics[width=1\linewidth]{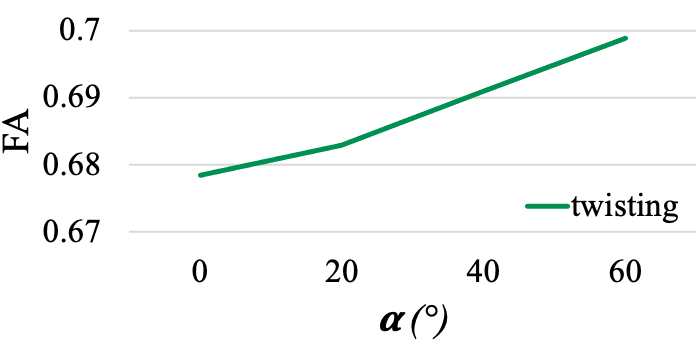}
    \caption{}
    \label{fig:FAtw}
    \end{subfigure}
    \hspace{1cm}
    \begin{subfigure}[normal]{.25\linewidth}
    \centering
    \includegraphics[width=1\linewidth]{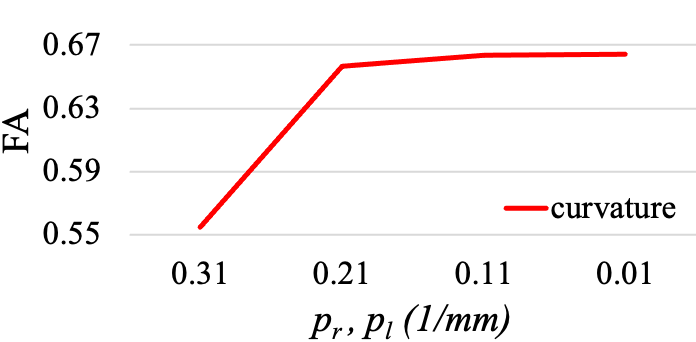}
    \caption{}
    \label{fig:FAcur}
    \end{subfigure}
\end{center}
\caption{Effect of a) increase in collagen density, b) increase in twisting, and c) decrease in curvature on FA.}
\label{fig:FACOllaTwCur}
\end{figure*}

\section{Discussion and Conclusion} \label{sec:Discussion}
\par This work aims to develop a workflow to generate a novel numerical phantom representing myocardial microstructure.
Compared with previous efforts, this study introduces two main novel contributions: (a) it considers more realistic shapes for the CMs and consequently a realistic complexity for the medium (relative to previous phantoms), by incorporating the native PDFs of CM shape parameters into the phantom;
(b) it models the ICDs within the phantom and takes into account their effects on the dMRI signal.
\par Interestingly, the comparison of the in-silico images with the histological images in Figure \ref{fig:VisualAssessment} shows that the proposed method closely mimics myocardial tissue.
The most striking observation emerges from the shape comparison of several single in-silico CMs with real CMs, shown in Figure \ref{fig:CMcomparison}, where the proposed algorithm is demonstrated to generate a realistic in-silico version of CMs.
Figure \ref{fig:CrossSection2} and Figure \ref{fig:LongCross} show an apparent similarity between the shape of the transverse and longitudinal cross-section of in-silico tissue and real ones, where combining both confirms an elliptical shape for CMs. 
However, closer inspection in Figure \ref{fig:CrossSection2} reveals different dispersion for crosswise orientation of the CMs (the directions of the principal axes of polygons) in in-silico images and histology. 
In the in-silico images, these directions are more correlated than those of the histological images.
\par The virtual morphometric study (Table \ref{t:MyoMeanVarMorphology}) confirms that the shape of individual CMs is consistent with real CMs, where \textit{p}-values for the length and volume of the CMs, i.e., \textit{p}$_{3}$, and \textit{p}-values for major and minor axes of the CMs, i.e., \textit{p}$_{2}$, are $>0.01$.
The reason for high \textit{p}-values for the length and volume lies in the step shown in Figures \ref{fig:SheetletAlgorithm}c and \ref{fig:SheetletAlgorithm}l-\ref{fig:SheetletAlgorithm}o, where the shape of each virtual CMs is modified to preserve the input PDF of the length and volume, respectively.
In contrast, the statistical tests related to PDFs of the major-axis and minor-axis result in lower values of \textit{p}$_2$.
As shown in Figure \ref{fig:MyoHistMorphology}, major-axis and minor-axis PDFs of the output are broadened than the input.
Broadening is a consequence of creating inter-CM space, during the transformation of the ellipse packing to the polygons. 
To create inter-CM space, the area of some polygons, shown in Figure \ref{fig:SheetletAlgorithm}.b, is shrunk, which leads to the reduction of the values of the major- and minor- axes and correspondingly, a broadening of their PDFs.

\par Considering microstructure complexity, the results of the structural correlator $\Gamma(k)$ corroborate the consistency between the in-silico generated tissue and real tissue (Figure \ref{fig:StructCooreCMs}).
\citep{novikov2014revealing} showed that the transverse cross-sections of skeletal muscle are classified as an extended disorder due to $\Gamma(k)\sim k^{-1}$.
This $k^{-1}$ behaviour comes from (relatively) straight lines of the myocyte's boundaries in the transverse cross-section of skeletal muscle, which spatially correlates over length scales of the cell's diameter.
As shown in Figure \ref{fig:CrossSection2}, CMs' boundaries in the transverse cross-section of the myocardial tissue are curved, and their directions are uncorrelated.
Therefore, the $k^{-1}$ gets cut-off at $k\sim \frac{1}{\textrm{Cell's}\quad\textrm{diameter}}$, and the disorder at larger cell diameter scales tends to plateau, i.e., $k^{0}$, the green dashed line in Figure \ref{fig:StructCooreCMs}, termed a short-ranged disorder.
\par This study shows that the median $\pm$ MAD of the angular distance between the input and simulated eigenvectors, along with the absolute angle difference between the input and simulated sheetlet angles, displayed in Figure \ref{fig:AngleDiffV} and \ref{fig:AngleDiffS}, are lower than those between DTI and STI, reported in Figure 10 and Figure DS3 of \citep{bernus2015comparison}.
Moreover, the angular distance between the input and simulated $V_3$ (Figure \ref{fig:V3}) is much smaller than the angular distance between the directly measured FLASH laminar normal and $V_3$ of STI and DTI, as reported in \citep{bernus2015comparison}.
Additionally, the resulting absolute angle difference for HA (Figure \ref{fig:AngleDiffS}) demonstrates that the deviation of the simulated HA from the input directions ($4.3^\circ\pm3.1^\circ$) is consistent with the absolute angle difference for HA between the experimental cDTI and histology reported by \citep{holmes2000direct} ($3.7^\circ\pm6.4^\circ$) and \citep{scollan1998histological} ($4.9^\circ\pm14.6^\circ$).
According to Figure \ref{fig:Box}, $\lambda_1$ and $\lambda_3$ are in reasonable agreement with those from the ex-vivo data. However, $\lambda_2$ of in-silico is considerably higher than its ex-vivo counterpart.
Since the ICS and ECS diffusivities contribute fairly well to the values of $\lambda_1$ and $\lambda_3$, the best way to reduce $\lambda_2$ with less effect on $\lambda_1$ and $\lambda_3$ is to add more obstacles to the passage of water molecules in the direction of $V_2$.
The most important compartments that hinder the movement of water molecules in this direction are microvasculature, fibroblast, and collagens (non-CM compartment) (with the VF of 7.7\%, 2.5\% \citep{greiner2018confocal}, and 2\% \citep{haddad2017novel}, respectively), as they are perpendicular to $V_2$, aligned along CMs \citep{greiner2018confocal}.
Moreover, inter-sheetlet space in real myocardial tissue is more tortuous than its in-silico counterpart, which results in $\lambda_2$ reduction. 
Appendix \ref{appendix:SimplePhantom} illustrates the effect of adding the above-mentioned obstacles in ECS on eigenvalues over simple phantoms. 
For these phantoms, adding the obstacles in ECS reduces all eigenvalues $\lambda_1$ and $\lambda_3$ reduce by the same amount, whereas the reduction in $\lambda_2$ is two times more than both $\lambda_1$ and $\lambda_3$.
Thus, the results in Appendix \ref{appendix:SimplePhantom} appear to support our argument that dismissing obstacles in ECS increases $\lambda_2$.  
In addition, although there is no statistical difference between the in-silico and ex-vivo values of $\lambda_1$ and $\lambda_3$, there are differences between the range of their quartiles and extremes as shown in the box plots of Figure \ref{fig:Box}.
The possible explanation for these differences is that the reported values for CM shape parameters are an average from a large part of a heart, whereas these values are used for every small in-silico voxel.

\section{Limitations and Future Works}
The most important limitation is that non-CM compartments, along with tortuous inter-sheetlet space, are not included into the in-silico phantom. 
Moreover, the lack of local information about the CM shape parameters for every voxel makes these findings less generalisable. 
Thirdly, as shown in Table 3, among 23 adjustable parameters of the numerical phantom, only eigenvectors of in-silico voxels are exactly matched with their ex-vivo counterparts. 
For the remaining 20 parameters, we either discarded these parameters (such as collagen’s diffusivity and relaxation ($D_{collagen}$ and $T_{2_{collagen}}$), tissue twisting ($\alpha$), wall curvature ($p_r$ and $p_l$), etc.) or used the values reported in the literature (such as CMs' dimension ($\textit{L}$, $\textit{V}$, $\textit{A}$ and $\textit{B}$), CMs' permeability ($\textit{k}_{Sarco.}$ and $\textit{k}_{ICDs}$), etc.) which are likely to differ from the parameters of the ex-vivo data under comparison.
Therefore, there is a definite need for an imaging method for intact hearts that enables us to reveal this information.
Imaging to this aim could be based on synchrotron radiation X-ray phase-contrast imaging, which has facilitated the investigation of myocardial tissue in detail, as recently shown by \citep{pierpaoli2010quantitative, shinohara2016three, kaneko2017intact}.
\par Sensitivity analysis of cDTI parameters or simulated diffusion signal to microstructure parameters listed in Table \ref{t:Simpara}, along with incorporating non-CM compartments into the present phantom, will be the subject of future works.
Moreover, the proposed micro-scale numerical phantom can be integrated into the XCAT phantom \citep{segars20104d} to generate a micro-structurally informed numerical phantom of a whole cardiac organ.
This opens up an opportunity for virtual imaging trials in cardiac dMRI through simulating 3D cardiac dMRI images \textemdash, which are micro-structurally informed\textemdash, together with including the effect of cardiac contraction and respiratory motion on dMRI images. 

\section*{Acknowledgements}
ML is an Early Career Researcher funded by the European Union’s Horizon 2020 Research and Innovation Programme under the Marie Skłodowska-Curie Innovative Training Network B-Q MINDED (H2020-MSCA-ITN-2017-764513). The Royal Academy of Engineering funds AFF (CiET1819\textbackslash 19).
Moreover, computations in this work were undertaken on ARC3 and ARC4, the High-Performance Computing facilities at the University of Leeds, UK, and on Amazon Web Services via our middle-ware platform MULTI-X (www.multix.org).
Finally, the authors would like to acknowledge the support of Prof D Novikov, providing the code and feedback on structural correlation analysis.
We acknowledge helpful discussions with Dr S Coelho in the early phases of this work.

\appendix 
\section{Mesh Analysis}\label{appendix:MeshAnalysis}
\begin{figure}[!ht]
\begin{center}
    \begin{subfigure}[normal]{.7\linewidth}
    \centering
    \includegraphics[width=1\linewidth]{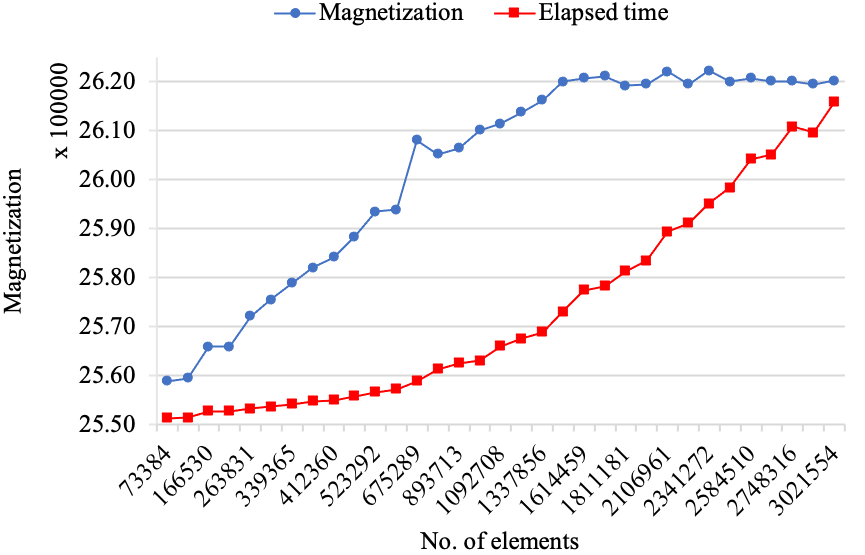}
    \caption{}
    \label{fig:MeshAnalysis-A}
    \end{subfigure}
    \begin{subfigure}[normal]{0.7\linewidth}
    \centering
    \includegraphics[width=1\linewidth]{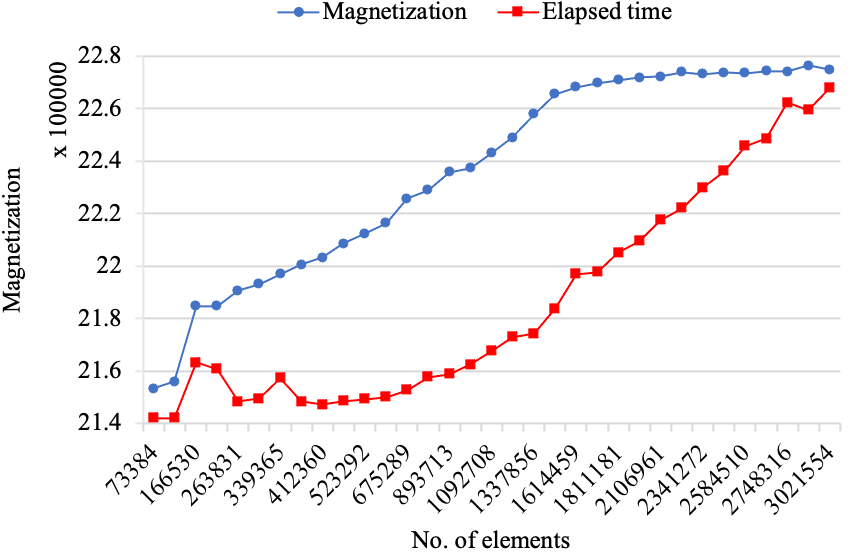}
    \caption{}
    \label{fig:MeshAnalysis-B}
    \end{subfigure}
\end{center}
\caption{Mesh Analysis: dependency of magnetisation to mesh resolution at: a) $b=0\:\ (\SI{}{\second\per\milli\meter^2})$ b) $b=100\:\ (\SI{}{\second\per\milli\meter^2})$}
\label{fig:meshanalysis}
\end{figure}

This experiment aims to find the coarsest tetrahedral mesh used to generate the phantom, at which the simulated signal becomes independent of the mesh resolution.
The mesh independence is evaluated by generating successively finer resolution of tetrahedral meshes for the domain of interest until the changes in the simulated signal become negligible.
The number of tetrahedral elements at which this mesh independency is observed is employed for all subsequent simulations.
Due to considerable memory requirements for higher b-values, we limited our analysis to the lower b-value of $b=0\ (\SI{}{\second\per\milli\meter^2})$ and $b=100\:\ (\SI{}{\second\per\milli\meter^2})$.
All imaging parameters of the simulation are set as described in Section 4.1, except diffusion encoding direction, which is set in the direction of z-axis.
Figure \ref{fig:meshanalysis} shows mesh analysis for a $100\times100\times100\:\ (\SI{}{\micro\meter^3})$ phantom where changes in magnetisation and elapsed time for solving Bloch\textemdash Torrey equation are evaluated versus the phantoms generated by more refined tetrahedral elements (or more tetrahedral elements) at $b=0\:\ (\SI{}{\second\per\milli\meter^2})$ and $b=100\:\ (\SI{}{\second\per\milli\meter^2})$.

According to Figure \ref{fig:meshanalysis}, the maximum number of tetrahedral elements at which the simulation result is approximately independent of mesh resolution is 1486993, which is achieved by setting the following parameters in iso2mesh:
\begin{itemize}
    \item opt.radbound = 6: The maximum surface element size.
    \item opt.angbound = 30: The minimum angle of a surface triangle.
    \item opt.distbound = 0.45: The maximum distance between the center of the surface bounding circle and centre of the element bounding sphere.
    \item opt.reratio = 3: The maximum radius-edge ratio.
    \item maxvol = 5: The target maximum tetrahedral element volume.
\end{itemize}

\section{Effect of adding obstacles into ECS on reduction in secondary eigenvalue} \label{appendix:SimplePhantom}
\begin{figure}[!ht]
\begin{center}
    \begin{subfigure}[normal]{.3\linewidth}
    \centering
    \includegraphics[width=1\linewidth]{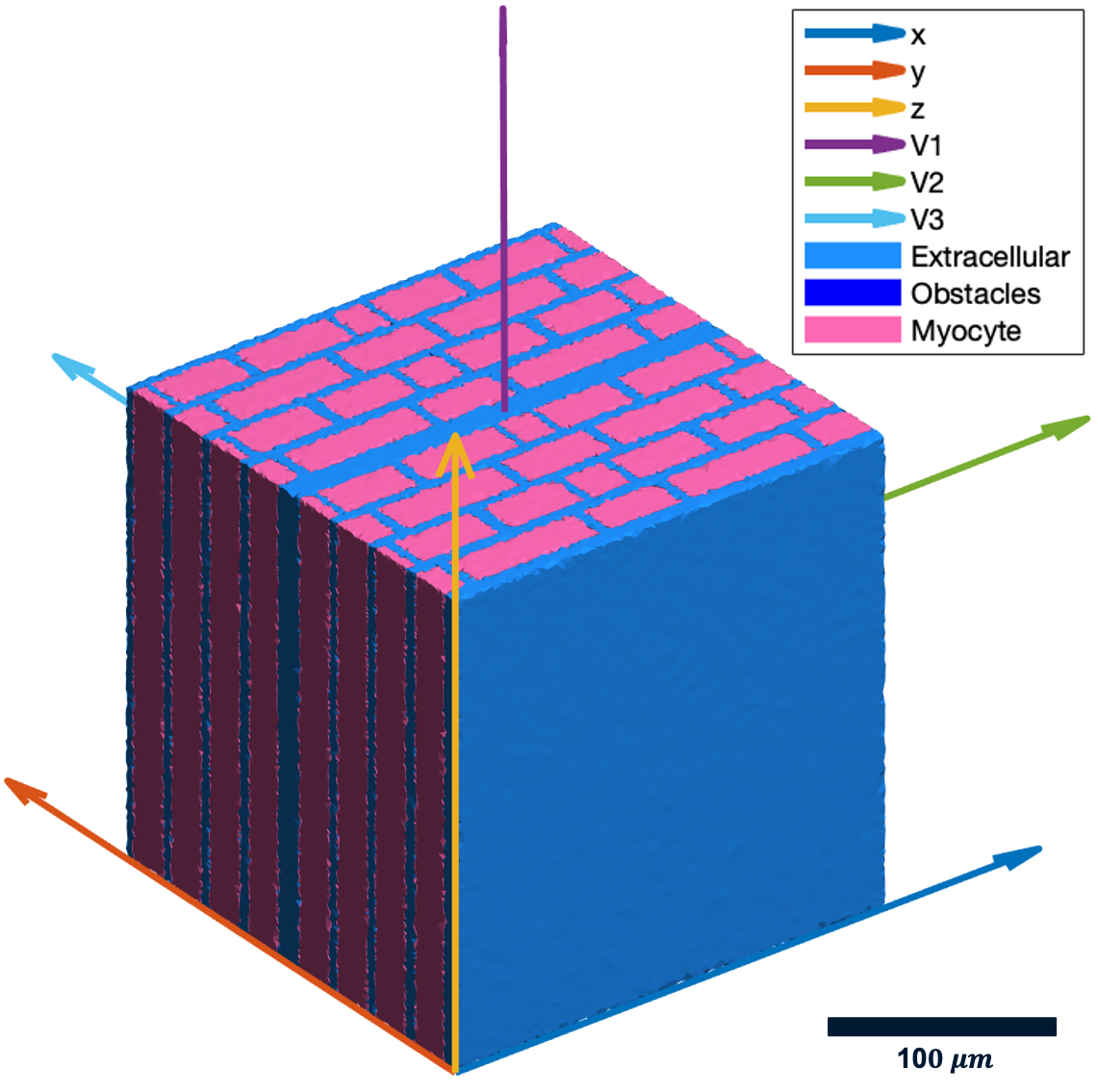}
    \caption{}
    \label{fig:SimplePhantoma}
    \end{subfigure}
    \begin{subfigure}[normal]{0.31\linewidth}
    \centering
    \includegraphics[width=1\linewidth]{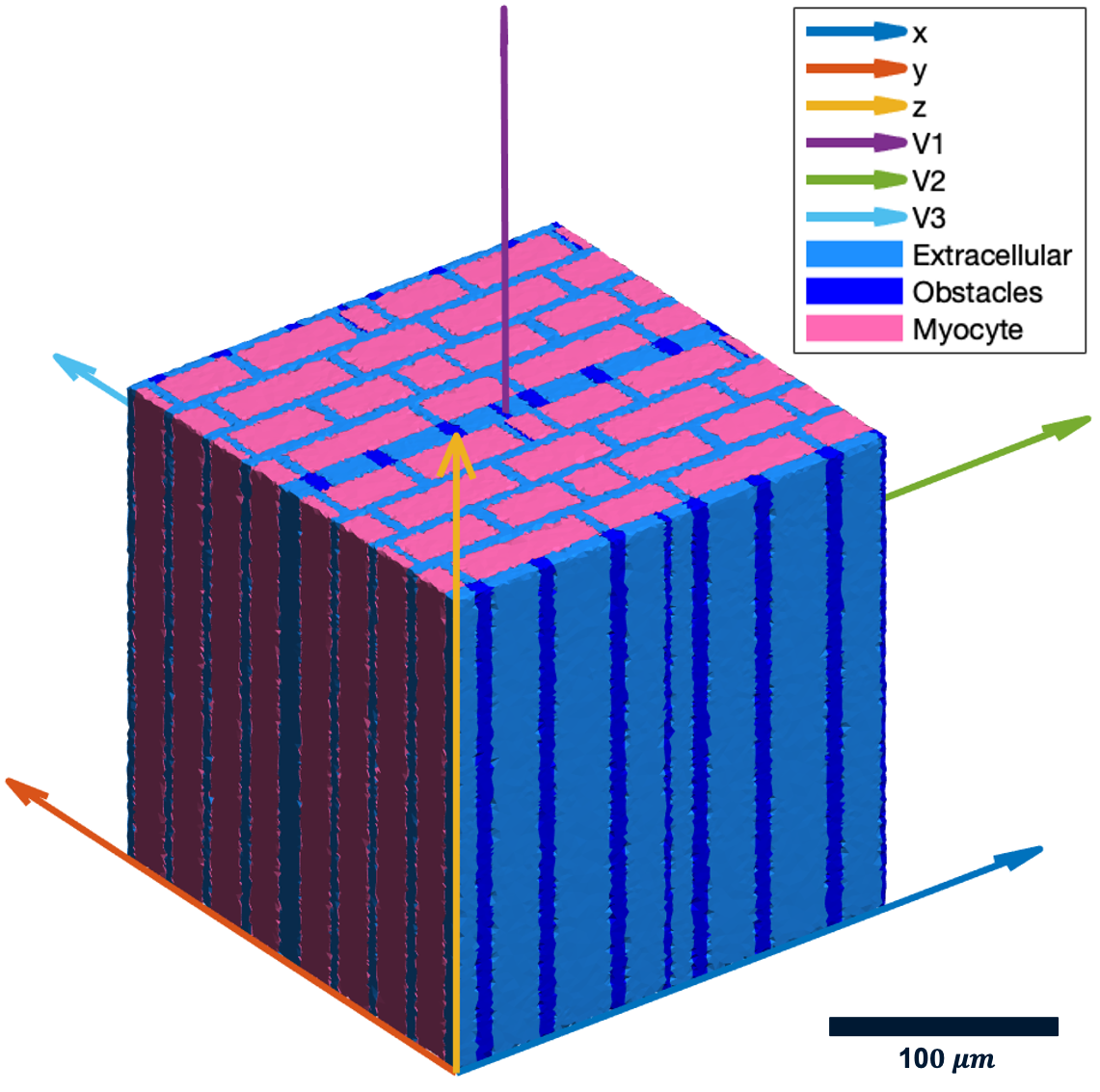}
    \caption{}
    \label{fig:SimplePhantomb}
    \end{subfigure}
    \begin{subfigure}[normal]{0.3\linewidth}
    \centering
\includegraphics[width=1\linewidth]{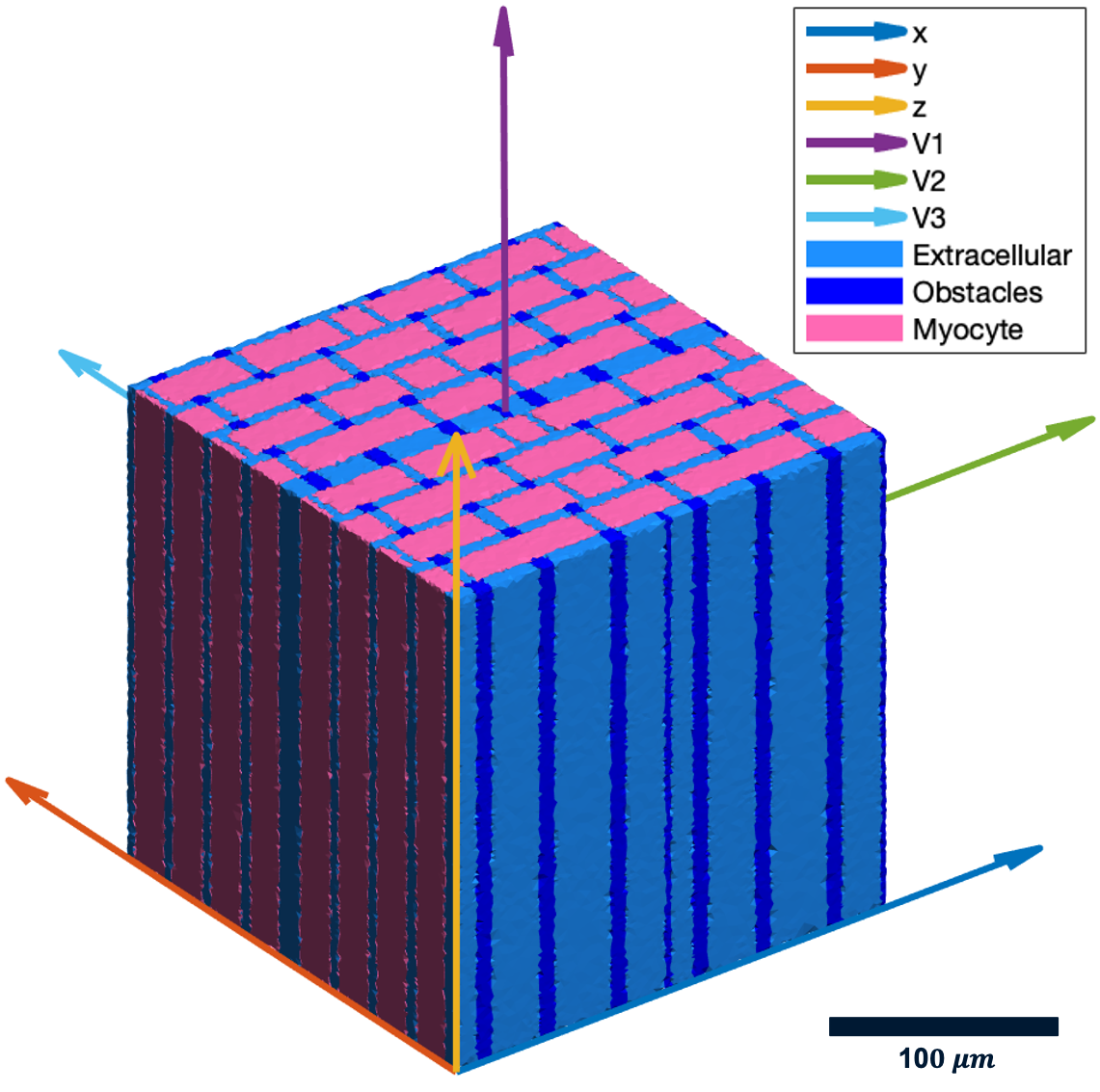}
    \caption{}
    \label{fig:SimplePhantomc}
    \end{subfigure}
    \begin{subfigure}[normal]{0.3\linewidth}
    \centering
    \includegraphics[width=1\linewidth]{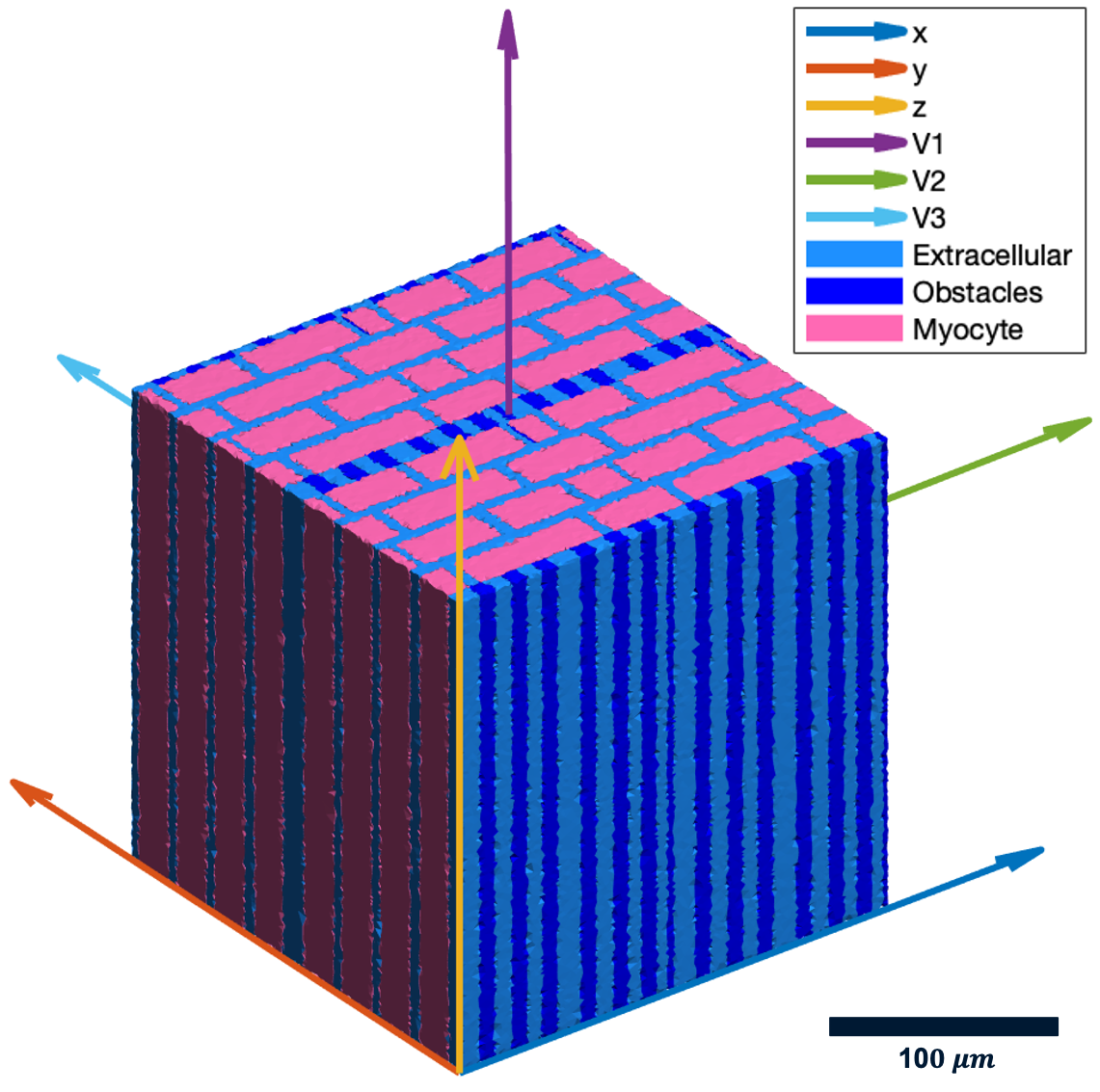}
    \caption{}
    \label{fig:SimplePhantomd}
    \end{subfigure}
    \begin{subfigure}[normal]{0.3\linewidth}
    \centering
    \includegraphics[width=1\linewidth]{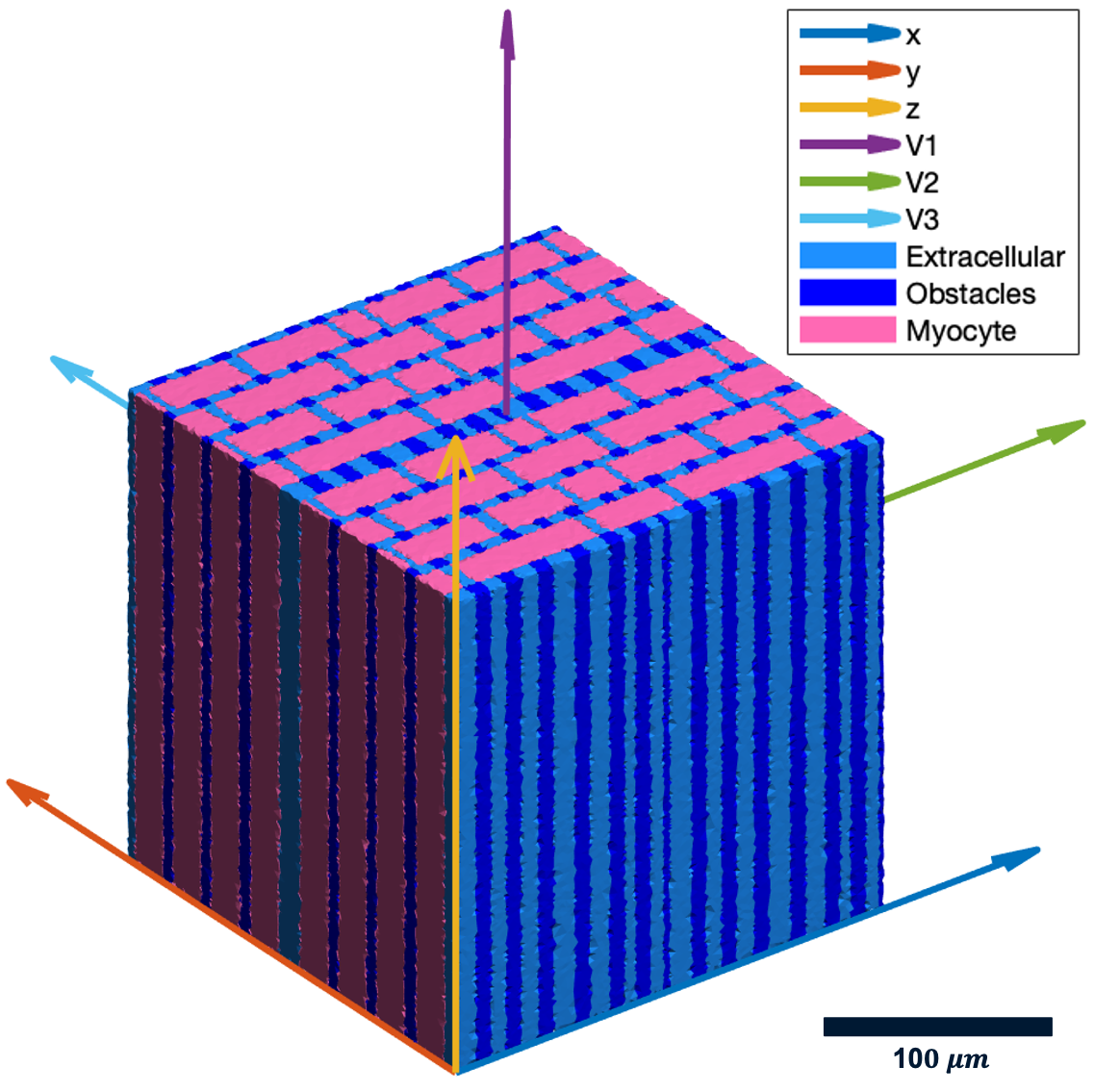}
    \caption{}
    \label{fig:SimplePhantome}
    \end{subfigure}
\end{center}
\caption{Simple phantom mimicking myocardial microstructure (light blue: ECS, pink: CMs) to investigate the effect of adding obstacles (dark blue) into the ECS on eigenvalues.}
\label{fig:SimplePhantom}
\end{figure}

To support the argument about the role of non-CM compartments and tortuous structure of inter-sheetlet space in reduction of $\lambda_{2}$, several simulations were run over multiple 3D simplified versions of myocardial in-silico phantoms with different density of the obstacles in ECS.
Figure \ref{fig:SimplePhantom} shows simplified phantoms where light blue is ECS, and pink and dark blue represent the CMs and obstacles, respectively. 
In these phantoms, both CMs and obstacles have cuboid shapes.
The phantom oriented in space, where the CMs and obstacles were parallel to the \textbf{z}-axis, and sheetlets were perpendicular to the \textbf{y}-axis.
In Figure \ref{fig:SimplePhantomb} and \ref{fig:SimplePhantomd} obstacles are only placed in inter-sheetlet space, whereas in Figure \ref{fig:SimplePhantomc} and \ref{fig:SimplePhantome} obstacles are also added into inter-CM space. 

In these experiments, parameters of the phantoms and simulations were similar to the experiment in Section 4, but spins were encoded in 12 directions.
Moreover, CMs and obstacles share the same values for diffusivity, $T_2$ relaxation, and permeability.
As illustrated in Figure \ref{fig:SimplePhantom}, the resulting eigenvectors for each phantom are oriented correctly, and adding the obstacles does not change their orientation, at least in the case of these simplified phantoms. 
Table \ref{t:SimplePhaEigenvelues1} shows eigenvalues for the phantoms in Figure \ref{fig:SimplePhantom} along with their percentage change with respect to the initial phantom in Figure \ref{fig:SimplePhantoma}, where there are no obstacles in ECS.
In all phantoms, the eigenvalues reduce, and the reduction in $\lambda_{2}$ is two times larger than the reduction in $\lambda_{1}$ and $\lambda_{3}$. 
Interestingly, in Table \ref{t:SimplePhaEigenvelues1} we observed that the reduction in both $\lambda_{1}$ and $\lambda_{3}$ is the same for all phantoms, e.g., $\Delta\lambda_{1}=\Delta\lambda_{3}$. Therefore, it is possible to hypothesise that these conditions are more likely to occur for the proposed in-silico phantom by adding the obstacles.

\begin{table*}[t]
\caption{Effect of adding obstacles into ECS (Figure 16) on reduction of $\lambda_{2}$ with ICS diffusivity of $\SI{0.83}{\micro\meter^2\per\milli\second}$.}
    \centering
    \begin{tabular}{ccccccc}
    \hline\hline
    Eigenvalues ($\SI{}{\micro\meter^2\per\milli\second}$) & Figure \ref{fig:SimplePhantoma} & Figure \ref{fig:SimplePhantomb} & Figure \ref{fig:SimplePhantomc} & Figure \ref{fig:SimplePhantomd} & Figure \ref{fig:SimplePhantome} \\
    \hline
    $\lambda_{1}$ & 1.63 & 1.56 ($\downarrow$4\%) & 1.47 ($\downarrow$10\%) & 1.51 ($\downarrow$7\%) & 1.35 ($\downarrow$17\%) \\
    $\lambda_{2}$ &	1.34 & 1.20 ($\downarrow$11\%) & 1.06 ($\downarrow$21\%) & 1.10 ($\downarrow$18\%) & 0.86 ($\downarrow$35\%) \\
    $\lambda_{3}$ &	1.01 & 0.97 ($\downarrow$4\%) & 0.90 ($\downarrow$10\%) &	0.94 ($\downarrow$7\%) & 0.83 ($\downarrow$18\%) \\
    \hline\hline
    \end{tabular}
\label{t:SimplePhaEigenvelues1}
\end{table*}

However, as mentioned earlier, including the above-mentioned obstacles reduces all eigenvalues, whereas we are only interested in $\lambda_{2}$ reduction.
Therefore, to preserve matching between $\lambda_{1}$ and $\lambda_{3}$, ICS diffusivity should increase to offset the reduction in $\lambda_{1}$ and $\lambda_{3}$, while more reduction in $\lambda_{2}$ is likely to lead to matching between $\lambda_{2}$ of in-silico and ex-vivo data.
This can be demonstrated by comparing $\lambda_{1}$ and $\lambda_{3}$ of phantoms (c) and (d) computed by ICS diffusivity of $\SI{1}{\micro\meter^2\per\milli\second}$, shown in Table \ref{t:SimplePhaEigenvelues2}, with their counterpart for phantom (a) computed using ICS diffusivity of $\SI{0.83}{\micro\meter^2\per\milli\second}$, shown in Table \ref{t:SimplePhaEigenvelues1}.

\begin{table*}[b]
\caption{Effect of adding obstacles into ECS (Figure \ref{fig:SimplePhantom}) on eigenvalues with ICS diffusivity of $\SI{1}{\micro\meter^2\per\milli\second}$.}
    \centering
    \begin{tabular}{ccccccc}
    \hline\hline
    Eigenvalues ($\SI{}{\micro\meter^2\per\milli\second}$) & Figure \ref{fig:SimplePhantoma} & Figure \ref{fig:SimplePhantomb} & Figure \ref{fig:SimplePhantomc} & Figure \ref{fig:SimplePhantomd} & Figure \ref{fig:SimplePhantome} \\
    \hline
    $\lambda_{1}$ & 1.81 & 1.73 & 1.63 & 1.69 & 1.52 \\
    $\lambda_{2}$ &	1.48 & 1.30 & 1.16 & 1.20 & 0.94 \\
    $\lambda_{3}$ &	1.08 & 1.04 & 0.97 & 1.01 & 0.91 \\
    \hline\hline
    \end{tabular}
\label{t:SimplePhaEigenvelues2}
\end{table*}


\bibliographystyle{model2-names.bst}\biboptions{authoryear}
\bibliography{main-manuscript}

\end{document}